\documentclass[aip,twocolumn,letterpaper]{revtex4}

\pdfoutput=1  

\usepackage{fullpage}

\usepackage{xcolor}

\usepackage{graphics}
\usepackage{epsfig}

\begin{document}
\title{Magnetic reconnection and thermal equilibration}
\author{Allen Boozer}
\affiliation{Columbia University, New York, NY  10027\\ ahb17@columbia.edu}

\begin{abstract}

When a magnetic field is forced to evolve on a time scale $\tau_{ev}$, as by footpoint motions driving the solar corona or non-axisymmetric instabilities in tokamaks, the magnetic field lines undergo large-scale changes in topology on a time scale approximately an order of magnitude longer than $\tau_{ev}$.  But, the physics that allows such changes operates on a time scale eight or more orders of magnitude slower.   An analogous phenomenon occurs in air.  Temperature equilibration occurs on a time scale approximately an order of magnitude longer than it takes air to cross a room, $\tau_{ev}$, although the physical mechanism that allows temperature equilibration is approximately four orders of magnitude slower than $\tau_{ev}$.  The use of Lagrangian coordinates allows the fundamental equations to be solved and both phenomena explained.  The paradigms and presumptions of traditional theories of magnetic reconnection are so ingrained that the understanding gained from analyses using Lagrangian coordinates has been largely ignored.   The theories of thermal equilibration and magnetic reconnection are developed in parallel to help readers obtain an understanding of the importance and implications of analyses using Lagrangian coordinates.

\end{abstract}

\date{\today} 
\maketitle

\section{Introduction}

A search on the Web of Science for the term \emph{magnetic reconnection} yields over fifteen thousand results.   The first is the 1956 paper of Parker and Krook \cite{Parker-Krook:1956}, which discusses the ``\emph{severing and reconnection of lines of force}" in connection with the magnetic dynamo problem.  This will be taken to be the definition of the term magnetic reconnection.  

A magnetic evolution in which the magnetic field lines can be interpreted as moving with a velocity $\vec{u}$ without severing and reconnecting will be called ideal.  In 1958, Newcomb \cite{Newcomb} showed that a magnetic evolution is ideal when
\begin{equation}
\frac{\partial\vec{B}}{\partial t} = \vec{\nabla}\times(\vec{u}\times\vec{B}). \label{ideal ev B}
\end{equation}
The velocity $\vec{u}$ of a magnetic field line along itself can be chosen for mathematical convenience; only the perpendicular velocity $\vec{u}_\bot$ is physically relevant.


In 1988, Schindler, Hesse, and Birn \cite{Schindler:1988} pointed out the fundamental paradox of fast magnetic reconnection.  The spatial scale $\Delta_d$ at which magnetic field lines cannot be distinguished because of the diffusivity of resistivity, $\eta/\mu_0$, is tiny compared to the scale $L$ at which reconnection is observed to rapidly occur.  In their description, the reconnection must be localized to a region, $\Delta_d$, in which the  local magnetic Reynolds number $(\mu_0 u_\bot/\eta)\Delta_d$ is unity.  The actual magnetic Reynolds number is defined globally;
\begin{eqnarray}
R_m &\equiv&\frac{\mbox{resistive time scale}}{\mbox{ideal evolution time scale}} \\ \nonumber\\
&=& \frac{\mu_0L^2/\eta}{L/u_\bot} \\ \nonumber\\
&=& \frac{\mu_0 u_\bot}{\eta} L.
\end{eqnarray}
In many important cases, $R_m$ is greater than $10^8$.  How can it be possible that the spatial scale at which distinguishability is lost, $\Delta_d$, can be so tiny compared to the scale over which rapid reconnection is observed, $L=R_m\Delta_d$?

There are two resolutions to the reconnection paradox.  The first resolution is to assume the plasma current is concentrated in layers of thickness $\Delta_d$, in which the current density $j\sim B_{rec}/\mu_0\Delta_d$, where $B_{rec}$ is the part of the magnetic field undergoing reconnection.  This current density is a factor of $R_m$ larger than the characteristic current density $B_{rec}/\mu_0L$.  This resolution is the only one considered by Schindler et al and has served as the fundamental paradigm for reconnection studies for the last sixty years.  Indeed, the formation and the maintenance of what is essentially a singular current density is often considered the fundamental problem in reconnection theory.   As will be shown, the an ideal evolution characteristically creates thin sheets of current but the maximum current density only rises linearly in time, which is too slow to explain observations.

The enhancement of the current density by a factor of $R_m$ during an ideal evolution appears to be even a greater problem than its maintenance.  Nonetheless, even maintenance is challenging, and plasmoid theory was developed \cite{Loureiro:2016} to address this issue.

The second resolution of the reconnection paradox is for some magnetic field lines that are separated by distance $\Delta_d$ at one point on their trajectories to be separated by a distance comparable to $L$ at another.  Magnetic field lines are defined at an instant in time.  The ratio of the maximum separation $\Delta_{max}$ between two lines that have a closest approach $\Delta_d$ changes over time in an ideal evolution.  Characteristically, $\Delta_{max}/\Delta_d$ is exponentially dependent on time divided by the characteristic evolution time, $\tau_{ev}=L/u_\bot$.  Reconnection becomes inevitable over the spatial scale $L$ when $\Delta_{max}/\Delta_d\sim R_m$, which occurs on a time scale of order $(\ln{R_m})\tau_{ev}$.  As will be shown, the required current density is only $\ln{R_m}$ larger than the characteristic current rather than $R_m$ times larger as in the first resolution.  The second resolution also incorporates reconnection due to electron inertia.  As discussed in  Section \ref{sec:Ohm's law} electron inertia gives a maximum distance at which evolving magnetic field lines can be distinguished, $\Delta_d=c/\omega_{pe}$.

Although the time scale required for the initiation of a fast reconnection, $(\ln{R_m}) \tau_{ev}$ is reminiscent of Petchek's slow-shock explanation for fast reconnection  \cite{Petschek}, the physics has little in common.

In 1973 Parker \cite{Parker:1973} noted that the rate of reconnection is often observed to be $\sim0.1V_A$, where $V_A$ is the Alfv\'en speed.  This has an obvious explanation.  Once a state is reached in which a significant fraction of the magnetic field lines have $\Delta_{max}/\Delta_d\sim R_m$, large-scale reconnection inevitably occurs, and static force balance is lost.  The unbalanced forces relax by Alfv\'en waves, both along and across the magnetic field lines.  Although the magnetic evolution due to Alfv\'en waves is ideal, Equation (\ref{ideal ev B}), the ratio of $\Delta_{max}/\Delta_d$ is naturally increases, as with any ideal evolution, which broadens the reconnection region.

The departure from an ideal evolution of the magnetic field, Equation (\ref{ideal ev B}), has little to do with Hall terms in Ohm's law. This can be seen using the generalized Ohm's law of Schindler, Hesse, and Birn \cite{Schindler:1988}, $\vec{E}+\vec{v}\times\vec{B} =\vec{\mathcal{R}}$, which can be rewritten as \cite{Boozer:part.acc} as $\vec{E}+\vec{u}_\bot\times\vec{B} = -\vec{\nabla}\Phi + \mathcal{E}_{ni}\vec{\nabla}\ell$.  The distance along a magnetic field line is $\ell$, the potential  $\Phi$ is a well-behaved function of space and time, and $\mathcal{E}_{ni}$ is a spatial constant along each magnetic field line.  The existence of the rewritten form is essentially obvious.  First choose $\Phi$ and $\mathcal{E}_{ni}$ to balance  $\vec{B} \cdot\vec{E}$.  The remaining terms are perpendicular to $\vec{B}$ and can be balanced by $\vec{u}_\bot\times\vec{B}$.    Faraday's law implies the magnetic evolution is ideal when  $\mathcal{E}_{ni}=0$, whatever the behavior of $\vec{v}$ may be.  Section \ref{sec:Ohm's law} has additional discussion.  

A common situation is that the speed of the ideal evolution $\vec{u}_\bot$ is very slow compared to the Alfv\'en speed before reconnection occurs.   In this situation, plasma inertia is unimportant, and the plasma velocity $\vec{v}$ has no direct relevance to reconnection---just the magnetic field line velocity $\vec{u}_\bot$.  

Although the sixty-year paradigm of reconnection theory is at odds with the second resolution, a simple example clarifies its importance.  Boozer and Elder \cite{Boozer-Elder} show that in a model of the solar corona  that $\Delta_{max}/\Delta_d$ tends to naturally increase exponentially on the time scale of the ideal evolution, the current density does lie in thin but wide ribbons along the magnetic field, but the maximum current density increases only linearly in time.  Figures based on numerical solutions of the model illustrate how an ideal evolution naturally leads to a state in which reconnection is inevitable.  Of course, a model is a model, and a general theory is essential.  

Here, a far more general theory of reconnection than that in Boozer and Elder \cite{Boozer-Elder} is developed, which is shown to be analogous in physics and mathematics to the equilibration of the temperature in a room.  The ideal equation for a thermal evolution is $\partial T/\partial t +\vec{v}\cdot\vec{\nabla}T =0$, with $\vec{\nabla}\cdot\vec{v}=0$, allows no change in the topology of the hot and cold regions.  An ideal evolution conserves the volume-averaged root-mean-square (RMS) deviation of the temperature from its average value; $\big<(T-\bar{T})^2\big>$, where $\bar{T}\equiv \big<T\big>$.   But in an actual room, $\big<(T-\bar{T})^2\big>$ relaxes on a time scale $\tau_r$, which is only an order of magnitude longer than  the time required for the flow of the air $\vec{v}$ to cross the room, the evolution time $\tau_{ev}\simeq L/v$.  Nevertheless,  thermal diffusion, which is required to break the topology of the the hot and cold regions and allow the relaxation of $\big<(T-\bar{T})^2\big>$, is intrinsically four orders of magnitude slower than $\tau_{ev}$.

Both magnetic reconnection and thermal relaxation are examples of mixing enhanced by orders of magnitude through stirring.  As defined in a major review \cite{Aref:2017}: \emph{``Stirring is advective redistribution, i.e., purely kinematic transport, and mixing is stirring together with diffusive effects."}   Both are applications of the mathematics of deterministic chaos and topological mixing.  A deterministic flow is chaotic when neighboring streamlines in a spatially bounded region have a separation that increases exponentially with time.  Articles on the mathematics of deterministic chaos and topological mixing can be easily found on the web, but the importance of these articles to this paper is only that such effects are common.

\begin{figure}
\centerline{ \includegraphics[width=2.5in]{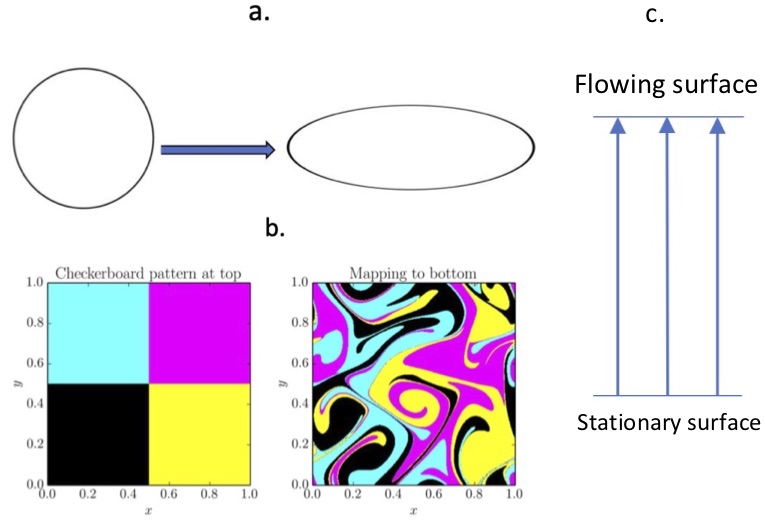}}
\caption{The same figure can be used to illustrate the evolution of tubes of streamlines of a time-dependent divergence-free velocity that depends on two coordinates and time, which has the form $\vec{v}_t =\hat{z}\times \vec{\nabla} h_t(x,y,t)$, or of a magnetic field at a fixed time, $\vec{B} = B_g\hat{z}+\hat{z}\times \vec{\nabla} h_b(x,y,z)$ with $B_g$ a constant guide field.  Figure \ref{fig:exp-fig}a shows how a tube of streamlines started on a circle will distort into an ellipse as time advances or a circular magnetic flux tube will distort into an ellipse as a function of $z$.  Figure \ref{fig:exp-fig}b illustrates how the tubes distort from squares if the streamlines are followed in a flow, which is periodic in $x$ and $y$, for a much longer in time or the magnetic field much further in $z$.   Figure \ref{fig:exp-fig}c illustrates a model \cite{Boozer:prevalence} that can be used to study reconnection when the initial magnetic field is a constant $B_g\hat{z}$.  The evolution is driven by a perfect conductor flowing with a velocity $\vec{v}_t=\hat{z}\times \vec{\nabla}h_t(x,y,t)$ at the  top of the diagram and a fixed perfect conductor at the bottom. Figure \ref{fig:exp-fig}b was drawn by Yi-Min Huang and used in A. H. Boozer, Nucl. Fusion \textbf{55}, 025001 (2015). }
\label{fig:exp-fig}
\end{figure}

The reason the time for topology breaking, $\tau_r$, is only an order of magnitude longer that the ideal-evolution advective time scale, $\tau_{ev}$, is that although ideal advection preserves topology it exponentially distorts shapes, Figure \ref{fig:exp-fig}.  Diffusion need only give transport across the thinest regions in these distorted shapes to destroy topology conservation, so diffusive effects enter the $\tau_r$ time scale logarithmically.   The natural logarithm of the large numbers that arise in physics lie within a factor of three of value twenty; $\ln(786)\approx20/3$ and $\ln(1.14\times10^{26})=3\times20$.  

To be convinced beyond what is possible by mathematics, construct diagrams of magnetic flux tubes as in Figure \ref{fig:exp-fig} and attempt to explain why $\eta/\mu_0$ diffusion across the exponentially thinning regions would not lead to rapid reconnection.   More detailed diagrams and discussions can be found in Boozer and Elder \cite{Boozer-Elder} for a model of magnetic reconnection in the solar corona.   

In astrophysics, a magnetic flux tube can imply that the magnetic field is far stronger within the tube than without.  Magnetic flux tubes are too important for understanding magnetic fields that vary smoothly in space to allow this implication to impede thought.  Here a magnetic flux tube is defined by the trajectories of a set of contiguous magnetic field lines.

Three concepts that are fundamental to the theory of both temperature equilibration and  magnetic reconnection will be discussed.

\begin{enumerate}
\item  Large forces arise unless the flow velocity is constrained so the flux of energy is equal to the energy density times the velocity.

The constraint for air flow is $\vec{\nabla}\cdot\vec{v}=0$.  The constraint on the velocity of magnetic field lines is $\vec{\nabla}\cdot\vec{u}_\bot+2\vec{u}_\bot\cdot\vec{\kappa}=0$, where $\vec{\kappa}$ is the field-line curvature.  

This constraint is obeyed for the same reason that a stream will flow down a narrow and circuitous ravine even when it could greatly shorten its path by flowing up and over the steep sides.  Two spatial coordinates are required to satisfy the $\vec{\nabla}\cdot\vec{v}=0$ constraint, but three are required to satisfy the analogous magnetic constraint.  Two-dimensional reconnection theory is unrepresentative just as one-dimensional thermal equilibration theory is.  


\item  The flow velocity must not only satisfy the energy-flux constraint but must also be chaotic.

Chaos does not mean the flow is indeterminate or turbulent.  The standard mathematical definition deterministic chaos is that the streamlines of a chaotic flow are in principle precisely calculable but have a separation that increases exponentially in time.  Chaotic flows preserve spatially bounded invariant surfaces $f(\vec{x},t)$ with $\partial f/\partial t+\vec{v}\cdot\vec{\nabla}f$, but $f$ becomes exponentially more convoluted and, therefore, sensitive to errors as time progresses, Figure \ref{fig:exp-fig}.  

Although the condition that the flow be chaotic may sound restrictive, it is a non-chaotic natural flow that is essentially impossible to realize \cite{Boozer-Elder,Aref:2017,Aref;1984} .  No special effort is required to achieve enhanced mixing by stirring.  Every cook knows that stirring enhances the mixing of fluids---no particular pattern of stirring or detailed computations are required.  

For fast magnetic reconnection, a chaotic flow $\vec{u}_\bot$ for the ideal evolution is required but is not sufficient.  A flow in two spatial dimensions $\vec{u}_\bot$ that is chaotic because of its temporal dependence will cause an exponential increase in the magnetic field strength, a case studied by Longcope and Strauss \cite{Longcope-Strauss:1994}.  Magnetic field lines are defined at a fixed time; their temporal dependence is irrelevant to whether they are chaotic at any instant of time.  A central question in toroidal fusion plasmas is whether magnetic surfaces exist, $\psi(\vec{x},t)$ with $\vec{B}\cdot\vec{\nabla}\psi=0$ with the constant-$\psi$ surfaces spatially bounded.  A chaotic ideal flow that makes the magnetic field lines chaotic does not directly break the magnetic surfaces, but the magnetic surfaces become exponentially more convoluted and sensitive to errors with distance along the magnetic field lines.


\item  Lagrangian coordinates, which are defined by the streamlines of the flow, allow exact solutions to be obtained even in chaotic flows.  

The importance of Lagrangian coordinates in this context was first explained in 1984 by Aref \cite{Aref;1984} who applied them to differential operators of the form $\partial T/\partial t +\vec{v}\cdot\vec{\nabla}T$.  In 1999, Tang and Boozer \cite{Tang-Boozer:1999} used Lagrangian coordinates to study the solutions to the full advection-diffusion equation, $\partial T/\partial t +\vec{v}\cdot\vec{\nabla}T=\vec{\nabla}\cdot (D_T\vec{\nabla}T)$.

In both thermal equilibration and magnetic reconnection, the ideal part of the evolution is intrinsically many orders of magnitude faster than the non-ideal, but without the non-ideal terms neither the constant-temperature contours nor the magnetic field lines can change their topology.  

\end{enumerate}


Before Aref's paper \cite{Aref;1984}, it was commonly assumed turbulence accounted for the enhancement of mixing by stirring.   Actually small-scale turbulence, $\ell_{turb}$,  impedes system-scale mixing as compared to advection-enhanced mixing with the same flow speed $v$.  Turbulence produces diffusive mixing with an effective diffusion coefficient $D_{turb}\approx \ell_{turb}\Delta v$, where $\Delta v$ is the change in the velocity over the scale $\ell_{turb}$, Equation (31.2) in \emph{Fluid Mechaninics} by Landau and Lifshitz \cite{LL-Fluid Mech}.  The time scale for turbulent mixing is $\tau_{turb}\approx L^2/D_{turb}$, so  $\tau_{turb}/\tau_{ev}\approx L/\ell_{turb}>>1$.  Optimal stirring has the same spatial scale as the desired region of mixing.  Even with optimal stirring, advective motion requires approximately ten times the evolution time $\tau_{ev}\approx L/v$ to produce the complex patterns that exponentially enhance the rate of mixing.  

A simple experiment that proves that turbulence is unnecessary for enhanced mixing is the preparation of a peanut butter and syrup sandwich---a skill mastered by American children shortly after they learn to pour their own glass of milk.  Put a blob of peanut butter and one of syrup on a plate and move a dinner knife back and forth through them.  Approximately ten strokes of the knife, they are throughly mixed---even the patience of a child is not taxed--but the viscosity is so high that neither substance moves except when pushed by the knife.

Turbulence can produce rapid reconnection as discussed in papers by a number of authors, for example Eyink, Lazarian, Matthaeus, and Vishniac \cite{Lazarian:1999, Eyink:2011,Eyink:2015,Matthaeus:2015,Lazarian:2020rev,Matthaeus:2020}.  Turbulent reconnection can be viewed as as example of the second resolution of the reconnection paradox.   Effects that are observed in an ideal evolution even with a smooth $\vec{u}_\bot$, such as as strong braiding of the plasma current \cite{Boozer-Elder}, have long been known in turbulent simulations.  Two points distinguish the theory of this paper and those on turbulent reconnection.  (1) Flows need not be turbulent to produce rapid reconnection---even smooth flows can.  (2) The most rapid reconnection for a given flow speed occurs when the gradients of the flow are comparable to the scale of reconnecting region. 

Lagrangian coordinates are rarely the optimal method for obtaining numerical solutions but provide constraints and conditions on the validity of solutions.  When topology breaking terms are ten to twenty orders of magnitude smaller than the dominant advective terms in the evolution, direct numerical simulations are not practical.  Methods based on Lagrangian coordinates rigorously determine the properties of solutions, but some find these methods unsettling.  A reviewer of an earlier paper felt that Lagrangian constraints on systems that cannot be solved numerically ``\emph{effectively makes his theory unfalsifiable by direct numerical simulations. Unfortunately, it also pushes his theory into the realm of non-science.}"  Actually, comparisons between numerical studies and Lagrangian-coordinate solutions are of great importance for determining when and how extrapolations can be made from what is calculable to what is needed to address practical problems.  

Analytic methods are at their most powerful in physics when the critical parameters are separated by many orders of magnitude.  Numerical methods are at their most powerful when all critical parameters are of a similar magnitude.  Methods based on Lagrangian coordinates are one example of this; adiabatic invariants are another.  The best known adiabatic invariant is the ratio of the energy of a dissipation-free pendulum divided by its frequency as the length of the pendulum is slowly changed.  As the change in length becomes slower, a numerical calculation of the amplitude of the swing becomes ever more inaccurate and time consuming, but the adiabatic invariant, which determines the amplitude, becomes more precisely conserved.

Lagrangian coordinates were applied to magnetic reconnection in several papers by Boozer in 2019 to (1) demonstrate  \cite{Boozer:ideal-ev} that the non-ideal part of the magnetic field grows exponentially in time, (2) show \cite{Boozer:part.acc}  that energetic particles can be accelerated even though the non-ideal parallel electric field is exponentially small, and (3)  determine \cite{Boozer:null-X} that magnetic field lines that pass within a distance of $c/\omega_{pe}$ of each other at any point on their trajectories cannot be distinguished in an evolution.  This changes the effect of magnetic nulls on reconnection theory \cite{Elder-Boozer}.  Magnetic helicity was shown to have only an exponentially small change during the reconnection process itself \cite{Boozer:part.acc}, but the spatial spreading of the parallel current caused by the reconnection can lead to a  rapid dissipation of the helicity during tokamak disruptions.  These papers showed the importance of Lagrangian coordinates to understanding magnetic reconnection, but appear to have had little impact within the reconnection community.  This presumably derives from fundamentally differences in method and conclusions from traditional studies.

New results on reconnection theory contained in this paper include: (1) the constraint that $\vec{\nabla}\cdot\vec{u}_\bot+2\vec{u}_\bot\cdot\vec{\kappa}=0$ to avoid changes in the magnetic-field energy, (2) the definition of the effective magnetic field, which simplifies the treatment of the $c/\omega_{pe}$ limit from electron inertia on field line distinguishability, (3) a clarified treatment of the current density, including a corrected expression for $j_{||}/B$. 

The objective in developing the theory of magnetic reconnection in parallel to the theory of thermal transport is to help readers obtain an understanding of the importance and nature of methods based on Lagrangian coordinates.  Neither magnetic reconnection nor thermal transport can be understood in the near-ideal limit without the use of this type of analysis.

Section  \ref{sec:T-equil}  derives the theory of temperature equilibration in a form that clarifies the physics and the mathematics of magnetic reconnection.  Section  \ref{sec:reconnection} applies the physics and mathematics developed in Section \ref{sec:T-equil} to magnetic reconnection.  Section \ref{sec:history} provides a summary and a context of this paper within the history of magnetic reconnection theory.


\section{Thermal equilibration \label{sec:T-equil} }

Magnetic reconnection has a simpler analogue in the establishment of thermal equilibrium in a room.  Both require a time scale comparable to an evolution time scale $\tau_{ev}$, which is defined by the gradient of a velocity,  the velocity of the air $\vec{v}$ or the magnetic field line velocity $\vec{u}$.  The thermal relaxation time or the reconnection time, $\tau_r$, is longer than the evolution time by approximately $\ln(\tau_D/\tau_r)$ where $\tau_D$ is the time scale that would be required for a diffusive relaxation, $\tau_D \equiv L^2/D$.  $L$ is a characteristic spatial scale and $D=D_T$, where $D_T$ is the thermal diffusivity of air, or $\eta/\mu_0$ for the resistive relaxation of a magnetic field.   In air $\tau_D/\tau_r\sim10^4$ but can be far larger in important cases of magnetic reconnection.

In thermal relaxation and magnetic reconnection, the equation for the evolution of the energy density, the thermal energy density $u=3nT/2$ or the magnetic energy density $B^2/2\mu_0$, constrains the form of the flow, $\vec{v}$ or $\vec{u}_\bot$.  What may be surprising is that the required number of spatial dimensions is also constrained for a fast relaxation, which means a relaxation on a time of order $\tau_{ev}$, two for thermal relaxation and three for magnetic reconnection.   

Proofs of enhancement require the use of Lagrangian coordinates of $\vec{v}$ or $\vec{u}_\bot$.  In thermal-type problems,  H. Aref \cite{Aref;1984} introduced Lagrangian coordinates in 1984 and demonstrated the importance of stirring; Tang and Boozer \cite{Tang-Boozer:1999} gave a complete solution in Lagrangian coordinates  in  1999.  Boozer \cite{Boozer:prevalence,Boozer:ideal-ev,Boozer:part.acc,Boozer:null-X} used Lagrangian coordinates to demonstrate the enhancement  of magnetic reconnection by the stirring that arises in the ideal evolution of a magnetic field in papers published in 2018 and 2019.

Heuristic arguments can and will be given for the exponential enhancement of mixing by stirring for two-dimensional thermal-equilibration problems. 


\subsection{Evolution  for thermal energy }

The constraint that $\vec{\nabla}\cdot\vec{v}=0$, for a slow flow to enhance thermal equilibration, arises from a thermodynamic relation, which says the flux of the thermal energy per unit volume $u$ is $(u+p)\vec{v}$ and not just $u\vec{v}$ as one would naively expect.  A similar constraint on the flow velocity of magnetic field lines follows from the flux of magnetic energy, $(B^2/\mu_0)\vec{u}_\bot$, where $B^2/\mu_0$ is the sum of the magnetic energy density and the magnetic pressure, both are $B^2/2\mu_0$, Section \ref{Sec:B-energy} and Equation (48) in \cite{Boozer:NA2015}. 

\subsubsection{Thermodynamic equation} 

Thermodynamics and mass conservation imply that in an entropy-conserving flow $\vec{v}$ that the internal energy of the fluid must evolve as $\partial u/\partial t + \vec{\nabla}(u\vec{v}) =- p\vec{\nabla}\cdot\vec{v}$, where $p$ is the pressure of the fluid, Equation (\ref{fluid energy}).  This is the fundamental constraint equation on thermal relaxation, and this section gives the derivation.

The standard thermodynamic relation $dU=TdS-pdV +\mu dN$ can be rewritten using energy, entropy, and particle densities, $U=uV$, $S=sV$, and $N=nV$.  Differentiation yields $V(du-Tds +\mu dn ) =- (u -Ts +p -n\mu)dV$, which implies that $du=Tds +\mu dn$ and that the chemical potential $\mu = (u -Ts +p)/n$ when the thermodynamic properties have no explicit dependence on the overall volume $V$ of the system.  Consequently,
\begin{eqnarray}
du&=& \frac{u+p}{n}dn + T(ds -\frac{s}{n} dn) \\
&=& \frac{u+p}{\rho}d\rho + \rho T ds_p. \label{thermo-relation}
\end{eqnarray}
where $\rho =m_pn$ is the mass density of particles with a mass $m_p$ and $s=s_p\rho$ with $s_p$ the entropy per unit mass---effectively the entropy per particle.

A corollary of Equation (\ref{thermo-relation}), called a Legendre transformation, is
\begin{eqnarray}
&&d\left\{\left(\frac{u+p}{\rho}\right)\rho\right\} = \rho d\left(\frac{u+p}{\rho}\right) + \left(\frac{u+p}{\rho}\right) d\rho. \hspace{0.2in}  \\
&& \mbox{Using Eq. (\ref{thermo-relation}),} \nonumber\\
&&d\left(\frac{u+p}{\rho}\right) = \frac{d p}{\rho} + T ds_p. \label{Legendre-p}
\end{eqnarray}

In addition to the thermodynamic equations, mass conservation implies
\begin{equation}
 \frac{\partial\rho}{\partial t} = - \vec{\nabla} \cdot(\rho \vec{v}). \label{continuity} 
 \end{equation}

The ideal thermodynamic relations, Equations (\ref{thermo-relation}) and (\ref{Legendre-p}), relate changes that can be spatial or temporal.  Equations (\ref{thermo-relation}) and (\ref{continuity}) imply
 \begin{eqnarray}
  &&\frac{\partial u}{\partial t} = \frac{u+p}{\rho}  \frac{\partial \rho}{\partial t} + \rho T \frac{\partial s_p}{\partial t}, \hspace{0.2in} \mbox{    and   }  \label{temporal thermo} \\
  &&\frac{\partial u}{\partial t} + \vec{\nabla}\cdot\{(u+p)\vec{v}\} = \rho\vec{v}\cdot\vec{\nabla} \left(\frac{u+p}{\rho}\right) \nonumber\\
  && \hspace{1.4in} +\rho T \frac{\partial s_p}{\partial t}.
  \end{eqnarray}
Writing the $d$'s as gradients in the Legendre-transformed Equation (\ref{Legendre-p}),
  \begin{eqnarray}
   \frac{\partial u}{\partial t} + \vec{\nabla}\cdot(u\vec{v}) &=& - p\vec{\nabla}\cdot\vec{v} +\rho T \left(\frac{ds_p}{dt}\right)_L \\
   &=& - p \vec{\nabla}\cdot \vec{v},  \label{fluid energy} 
  \end{eqnarray}
  when $(\partial s_p/\partial t)_L\equiv \partial s_p/\partial t+\vec{v}\cdot\vec{\nabla}s_p =0$ as is the case in an ideal evolution.  The Lagrangian derivative $(\partial s_p/\partial t)_L $ is the rate of change in the frame of the moving fluid.  The evolution of the entropy can also be written as
   \begin{eqnarray}
   &&\rho\left(\frac{\partial s_p}{\partial t}+\vec{v}\cdot\vec{\nabla}s_p\right)=\frac{\partial s}{\partial t} + \vec{\nabla}\cdot(s\vec{v}). \label{alt-s}
   \end{eqnarray}
   

\subsubsection{Implications for temperature evolution}

 Equation (\ref{fluid energy}) implies that a large change in the energy density occurs when $\vec{\nabla}\cdot \vec{v} \neq 0$.  Mixing can be enhanced by slow stirring only when $\vec{\nabla}\cdot \vec{v}= 0$.  A divergent velocity gives sound waves, Appendix \ref{sec:v-ev}.  The curl of the velocity, the vorticity $\vec{\nabla}\times \vec{v}$, is driven by a temperature gradient crossed with gravitational acceleration $\vec{g}$, Appendix \ref{sec:v-ev}.
 
 When $\vec{\nabla}\cdot \vec{v}=0$, Equation (\ref{fluid energy}) implies 
 \begin{eqnarray}
  &&\frac{\partial u}{\partial t} + \vec{v}\cdot\vec{\nabla}u = 0 \mbox{ and } \label{div-free}\\
&&\frac{\partial T}{\partial t} + \vec{v}\cdot\vec{\nabla}T = 0.  \label{ideal T} 
 \end{eqnarray}
 using $u= (3/2) \rho T/m_p$ in a monatomic ideal gas, where $m_p$ is the mass of each particle, and mass conservation, Equation (\ref{continuity}).

The volume averaged of squared temperature variation cannot change in an ideal evolution with $\vec{\nabla}\cdot \vec{v}=0$ for then Equation (\ref{ideal T} ) implies
\begin{eqnarray}
&&\frac{\partial T^2}{\partial t} +\vec{\nabla}\cdot(T^2\vec{v}) = 0 \\
&& \frac{d}{dt}\Big< T^2 \Big> \equiv \frac{\int \frac{\partial T^2}{\partial t} d^3x}{\int d^3x} =0
\end{eqnarray}
Similarly, $d\bar{T}/dt=0$, where $\bar{T}=\big< T \big>$, so $\big< (T-\bar{T})^2 \big>$ cannot change when the temperature has an ideal evolution with $\vec{\nabla}\cdot \vec{v}=0$.

 
 \subsubsection{Full ideal energy conservation}
 
 The ideal evolution of the thermal energy is not commonly given in the form of Equation (\ref{fluid energy}) but in the full energy-conservation form, which is given to avoid confusion.
 
 The equation of motion of a gas subject to gravity, $\vec{g}=- \vec{\nabla}\Phi_g$, is
\begin{eqnarray}
&&\rho \frac{\partial \vec{v}}{\partial t} +\rho\vec{v}\cdot\vec{\nabla}\vec{v} = -\vec{\nabla}p - \rho \vec{\nabla}\Phi_g. \label{fluid-force}
\end{eqnarray}
This equation, mass conservation, Equation (\ref{continuity}), energy evolution, Equation (\ref{fluid energy}), and vector identities imply
 \begin{eqnarray}
&& \frac{\partial}{\partial t}\left(\frac{1}{2} \rho v^2 + u+\rho \Phi_g\right) \nonumber\\
&& \hspace{0.3in} + \vec{\nabla}\cdot \left\{\left(\frac{1}{2} \rho v^2 + u+p +\rho \Phi_g\right)\vec{v}\right\} =0, \hspace{0.2in} \label{gas energy}
\end{eqnarray}
which is the complete equation for ideal energy conservation in a gas.

 Since Equation (\ref{fluid energy}) implies the  volume average $\left< u \right>$ does not change in an ideal evolution with $\vec{\nabla}\cdot\vec{v}=0$, Equation (\ref{gas energy}) implies that $\left< \rho v^2/2  +\rho \Phi_g \right>$ cannot change either.  Since $\Phi_g=gy$ with $y$ the vertical Cartesian coordinate, the pressure changes little over the volume the size of a room, and $T=T_0+\tilde{T}$ with $\tilde{T}<<T_0$, one finds $\delta\left< v^2/2\right>\approx (g/T_0)\delta \left< y\tilde{T} \right>$.  When the initial $\tilde{T} = - (y/H) \delta \tilde{T}_0$, where $H$ is the height of the ceiling
\begin{equation}
\left< v^2\right> \approx \frac{\delta \tilde{T}_0}{T_0}gH, \label{exp.v^2}
\end{equation}
when the variation in $\tilde{T}$ with $y$ is removed.
 
 
 \subsubsection{Diffusive energy relaxation}
 
 Equation (\ref{ideal T} ) for the ideal evolution of the temperature with a divergent free flow preserves in some form the variations in the temperature.  Stirring produces extreme spatial complexity in $T-\bar{T}$ but cannot reduce its RMS amplitude or the topology of hot and cold regions.  A related issue arises in magnetic reconnection. Stirring due to the velocity $\vec{u}_\bot$ associated with an ideal magnetic evolution produces extreme spatial complexity in flux tubes formed by magnetic lines but cannot change which lines are in a particular tube.
 
The RMS amplitude of $T-\bar{T}$, does relax in the presence of diffusive energy transport, which modifies Equation (\ref{div-free}) to
\begin{equation}
 \frac{\partial u}{\partial t} + \vec{\nabla} \cdot(u\vec{v}+\vec{q}_d) = 0,  \mbox{   where   } \label{diffusive energy}
 \end{equation} 
 \begin{eqnarray}
 \vec{q}_d&=&- \frac{3}{2} D_T\vec{\nabla}T  \mbox{   with   } \\
 D_T &\approx& 2.2\times10^{-5}~\frac{\mbox{m}^2}{\mbox{s}} \mbox{  for air.  }
 \end{eqnarray}
 The relaxation of the RMS amplitude of $T-\bar{T}$ requires a time $\tau_r$, which is the evolution time $\tau_{ev}$ multiplied by a term that depends on $\ln(1/D_T)$ as $D_T\rightarrow0$.  The time scale for magnetic reconnection is also the ideal evolution time multiplied by a term with a logarithmic dependence on $\eta/\mu_0$.
 
The diffusive energy flux enters the thermodynamic derivation through the entropy.  Instead of being constant, $ds_p/dt=0$, the entropy per particle evolves as $\partial s_p/\partial t +\vec{v}\cdot\vec{\nabla}s_p = -(\vec{\nabla}\cdot \vec{q}_d)/(\rho T)$ as noted by Landau and Lifshitz in Equation (49.4) of \emph{Fluid Mechanics} \cite{LL-Fluid Mech}. Using Equation (\ref{alt-s}) for the evolution of the entropy per unit volume $s$ instead of the entropy per particle,
\begin{eqnarray}
&&\frac{\partial s}{\partial t} + \vec{\nabla}\cdot(s\vec{v})=- \frac{\vec{q}_d}{T};\\
&&\frac{dS}{dt}=-\int \frac{\vec{\nabla}\cdot\vec{q}_d}{T} d^3x \\
&&\hspace{0.25in} = - \int \frac{\vec{q}_d\cdot\vec{\nabla}T}{T^2} d^3x,
\end{eqnarray}
where the total entropy $S=\int s d^3x$.

 
 \subsubsection{Heuristic relaxation estimate}
 
  In a monatomic ideal gas, $u=3p/2$ and $p=\rho T/m_p$, Equation (\ref{diffusive energy}) implies the temperature relaxes as
 \begin{equation}
 \frac{\partial T}{\partial t} + \vec{v}\cdot\vec{\nabla}T = - \frac{2}{3}\vec{\nabla}\cdot\vec{q}_d, \label{T-ev}
 \end{equation}
 with $\vec{\nabla}\cdot\vec{v}=0$.  This is the advection-diffusion equation for the temperature.

 If $D_T=0$, a contour of constant temperature moves with the flow velocity $\vec{v}$.  To keep the argument simple, suppose $z$ is a symmetry direction, so the constant-$T$ contours and the $\vec{v}$ are in the $x-y$ plane.  Let $\vec{x}_e$ be the location of an extremum (maximum or minimum) of the temperature, then a constant-$T$ contour is a closed curve with points located at $\vec{x}=\vec{x}_e +\vec{\delta}$.  Consider a small $\vec{\delta}$, then fixed temperature points move as 
 \begin{eqnarray}
 \frac{d\vec{x}}{dt} &=& \vec{v}(\vec{x},t)\\
 &\simeq& \vec{v}(\vec{x}_e,t) + (\vec{\delta}\cdot\vec{\nabla})\vec{v} \\
\frac{d\vec{\delta}}{dt} &\simeq& \vec{\delta}\cdot\vec{\nabla}\vec{v}, \label{delta-ev}
 \end{eqnarray}
 an equation that has solutions that exponentiate in time.  
 
 When the original constant-$T$ contour is circular, the contour distorts into an ellipse, Figure \ref{fig:exp-fig}a, with radii $\delta_{max}$ and $\delta_{min}$. Incompressibility implies the area of the contour is fixed, so $\delta_{max}\delta_{min}$ is constant.  One of the two radii, $\delta_{max}$ grows exponentially on the evolution time scale,
 \begin{equation}
 \tau_{ev}\equiv \frac{1}{\Big| \vec{\nabla}\vec{v} \Big|},
 \end{equation}
where $\big| \vec{\nabla}\vec{v} \big|$ is the largest component in the $2\times2$ matrix $\vec{\nabla}\vec{v}$.  The other radius of the ellipse  $\delta_{min}$ shrinks with their product giving the square of the radius of the original circle.  

The behavior of a constant-$T$ contour changes fundamentally when the longer of the two radii of the ellipse $\delta_{max}$ reaches the spatial scale of the velocity variation.  Then, the constant-$T$ contours fold back on themselves and become extremely complicated but the narrowest places on the constant-$T$ contours $\delta_{min}$  continue to decrease approximately exponentially Figure \ref{fig:exp-fig}b.  That is, approximately as $e^{-\gamma_{ev}t/\tau_{ev}}$, where $\gamma_{ev}$ has a complicated spatial and temporal dependence, but $\gamma_{ev}=1/3$ can be representative.  

It is the folding back of the contours that slows the rate of thermal relaxation to quasi-diffusive when the spatial scale of the flow is far smaller than $L$ the scale of the room in which the relaxation takes place.

Diffusion is becomes faster quadratically as $\delta_{min}$ decreases, which causes the constant-$T$ contours to break on the characteristic time scale 
\begin{eqnarray}
\tau_r &=& \tau_D e^{-2\gamma_{ev}\tau_r/\tau_{ev}},  \mbox{     or    } \\
&=&\frac{\tau_{ev}}{2\gamma_{ev}} \ln\left(\frac{\tau_D}{\tau_r}\right),  \mbox{  where  }\\
\tau_D&=& \frac{L^2}{D_T}\\
&\approx& 1.1\times 10^6~\mbox{sec} \approx 13~\mbox{days}.
\end{eqnarray}
and $L\approx5~$m is the greatest distance through which the temperature must relax.
 
 The expected RMS velocity from energy conservation, Equation (\ref{exp.v^2}), gives an estimate of the evolution time.  When the ceiling height is 3~m and $\delta \tilde{T}_0/T_0=10^{-2}$,
 \begin{eqnarray}
 \tau_{ev} &\approx& \frac{2L}{\sqrt{\frac{\delta \tilde{T}_0}{T_0}gH} } \approx 18~\mbox{s}, \mbox{   so   }\\
 \tau_r &\approx& \frac{\tau_{ev}}{2\gamma_{ev}} \ln\left(\frac{\tau_D}{\tau_r}\right)\approx230~\mbox{s}
 \end{eqnarray}
 when $\gamma_{ev}=1/3$.  The finite-time Lyapunov exponent of the flow is $\gamma_{ev}/\tau_{ev}$.


\subsection{Lagrangian coordinates}

 Lagrangian coordinates are the method of characteristics when applied to differential operators of the form $\partial T/\partial t +\vec{v}\cdot\vec{\nabla}T$.  The standard problem is finding the evolution of $\vec{\nabla}T$ in the presence of both advection $\vec{v}\cdot\vec{\nabla}T$ and diffusion $\vec{\nabla}\cdot(D_T\vec{\nabla}T)$.  
 
 Lagrangian coordinates are more important for ascertaining the properties of solutions to the advection-diffusion equation, Eq. (\ref{T-ev}), for a non-zero $D_T$ as $D_T\rightarrow0$, than in obtaining explicit solutions.   Information on low diffusivity limit  is particularly important for magnetic reconnection in the solar corona, where the ratio of the advective to the resistive terms, the magnetic Reynolds number $R_m$, can reach $10^{12}$.  Direct numerical simulations become impractical in the low diffusivity limit, and Lagrangian coordinates are the only practical theoretical method of determining the properties of solutions. 


\subsubsection{Definition of Lagrangian coordinates}

Let $\vec{x}_0(x_0,y_0,z_0)$ give positions in space as functions of $x_0,y_0,z_0$ at $t=0$.  For example, Cartesian coordinates can be used to define positions, $\vec{x}_0=x_0\hat{x} +y_0\hat{y}+z_0\hat{z}$, but the set of three coordinates is essentially arbitrary.  The coordinates $x_0,y_0,z_0$ become Lagrangian coordinates when positions in space are defined by $\vec{x}(x_0,y_0,z_0,t)$, which can be more compactly written as $\vec{x}(\vec{x}_0,t)$, with
\begin{eqnarray}
\left(\frac{\partial \vec{x}}{\partial t}\right)_L  &\equiv& \frac{\partial\vec{x}(\vec{x}_0,t)}{\partial t} \\
&=& \vec{v}(\vec{x},t). \label{Langrangian-def}
\end{eqnarray}
Positions in space at any fixed point in time can be described by three coordinates $x,y,z$, and Equation (\ref{Langrangian-def}) determines the functions $x(x_0,y_0,z_0,t)$, $y(x_0,y_0,z_0,t)$, and $z(x_0,y_0,z_0,t)$, which give the location of a point at time $t$ that was at $(x_0,y_0,z_0)$ at $t=0$. 
\begin{eqnarray}
\left(\frac{\partial T}{\partial t}\right)_L &=&\left(\frac{\partial T}{\partial t}\right)_{\vec{x}} + \frac{\partial T}{\partial\vec{x}} \cdot \left(\frac{\partial \vec{x}}{\partial t}\right)_L \nonumber\\
&=&\left(\frac{\partial T}{\partial t}\right)_{\vec{x}} + \vec{v}\cdot\vec{\nabla} T.
\end{eqnarray}

\subsubsection{Evolution of the gradient of a function}

When a function $T(\vec{x},t)$ is carried by a flow, as is the temperature when   $\partial T/\partial t +\vec{v}\cdot\vec{\nabla}T=0$, then an important question is how does $\vec{\nabla}T$ evolve given the initial gradient in the function, $\vec{\nabla}_0 T$;
\begin{eqnarray}
\vec{\nabla}_0 T &=& \frac{\partial T}{\partial \vec{x}_0} \\
&=&  \frac{\partial T}{\partial \vec{x}}\cdot\frac{\partial \vec{x} }{\partial \vec{x}_0}\\
&=& \vec{\nabla} T \cdot\tensor{J}; \label{Lag-grad}\\
\tensor{J} &\equiv&\frac{\partial \vec{x} }{\partial \vec{x}_0}\\
&=& \left(\begin{array}{ccc}\frac{\partial x}{\partial x_0} &\frac{\partial x}{\partial y_0} &\frac{\partial x}{\partial z_0} \\ \frac{\partial y}{\partial x_0} &\frac{\partial y}{\partial y_0} &\frac{\partial y}{\partial z_0}  \\ \frac{\partial z}{\partial x_0} &\frac{\partial z}{\partial y_0} &\frac{\partial z}{\partial z_0} \end{array}\right).
\end{eqnarray}

These algebraic manipulations may be more obvious using coordinate components with the three Lagrangian coordinates numbered by Greek superscripts $x_0^\alpha$ and the three ordinary spatial coordinates numbered by Latin superscripts $x^i$, then
\begin{eqnarray}
\frac{\partial T}{\partial x_0^\alpha} &=& \sum_i \frac{\partial T}{\partial x^i} \frac{\partial x^i}{\partial x_0^\alpha};\\
J^i_{\hspace{0.05in}\alpha} &\equiv& \frac{\partial x^i}{\partial x_0^\alpha}.
\end{eqnarray}

The Jacobian matrix $\tensor{J}$ can be written in the Singular-Value-Decomposition (SVD) form,
\begin{eqnarray}
\tensor{J}&=& \tensor{U}\cdot\left(\begin{array}{ccc}\Lambda_u & 0 & 0 \\0 & \Lambda_m & 0 \\0 & 0 & \Lambda_s\end{array}\right)\cdot\tensor{u}^\dag \label{SVD} \nonumber\\
&=& \hat{U}\Lambda_u\hat{u}^\dag+ \hat{M}\Lambda_m\hat{m}^\dag+ \hat{S}\Lambda_s\hat{s}^\dag; \hspace{0.2in}  \label{comp SVD}\\
\hat{U} &=& \hat{M}\times\hat{S} \hspace{0.1in}\mbox{and}\hspace{0.1in} \hat{u} = \hat{m}\times\hat{s}.
\end{eqnarray}
Both $\tensor{U}$ and $\tensor{u}$ are unitary matrices, $\tensor{U}\cdot \tensor{U}^\dag=\tensor{1}$.  The three columns of $\tensor{U}$, which are $\hat{U}$, $\hat{M}$, and $\hat{S}$, are orthonormal unit matrix vectors $\hat{U}\cdot \hat{U}^\dag=1$ and $\hat{U}\cdot \hat{M}^\dag=0$.  Analogous relations exists between the three columns of  $\tensor{u}$, which are $\hat{u}$, $\hat{m}$, and $\hat{s}$.  The adjoint of a matrix column vector is a matrix row vector with the same three entrees.

The relation $\vec{\nabla}_0 T= \vec{\nabla} T \cdot\tensor{J}$ is equivalent to $\vec{\nabla}_0 T= \tensor{J}^\dag\cdot \vec{\nabla} T$.  Multiplying this relation by the left-inverse of $\tensor{J}^\dag$ gives
\begin{eqnarray}
\vec{\nabla}T &=& \left(\tensor{J}^{-1}\right)^\dag\cdot\vec{\nabla}_0T \nonumber\\
&=& \frac{ \hat{u}^\dag\cdot \vec{\nabla}_0T}{\Lambda_u}\hat{U} +\frac{ \hat{m}^\dag\cdot \vec{\nabla}_0T}{\Lambda_m}\hat{M}+\frac{ \hat{s}^\dag\cdot \vec{\nabla}_0T}{\Lambda_s}\hat{S}; \label{grad T} \nonumber\\
\end{eqnarray} 
\begin{equation}
\left(\tensor{J}^{-1}\right)^\dag =\frac{ \hat{U} \hat{u}^\dag}{\Lambda_u} +\frac{\hat{M} \hat{m}^\dag}{\Lambda_m}+\frac{\hat{S} \hat{s}^\dag}{\Lambda_s}.
\end{equation}

When the magnitude of the exponentiation is large, it can be calculated far more easily and accurately using the Frobenius norm of the Jacobian matrix than by an SVD.  The Frobenius norm $\big\| \partial \vec{x}/\partial \vec{x}_0 \big\|$ is the square root of the sum of the squares of the matrix elements and is also equal to the square root of the sum the squares of the singular values.  Exponentiation is of practical importance only when $\Lambda_u>>1$; in this limit $\big\| \partial \vec{x}/\partial \vec{x}_0 \big\|\rightarrow\Lambda_u$.  For these reasons, the Frobenius norm was used to measure exponentiation in reconnection example of Boozer and Elder \cite{Boozer-Elder}. 


\subsubsection{The Lagrangian Jacobian and the singular-value magnitudes}

The theorem that the determinant of a product of matrices is the product of the determinants implies that the Jacobian of  Lagrangian coordinates, which is the determinant of $\tensor{J}$, is
\begin{equation}
\mathcal{J}_L = \Lambda_u\Lambda_m \Lambda_s.
\end{equation}

The time derivative of the Jacobian can be determined by writing mass conservation in the form $(\partial \rho/\partial t)_L =-\rho \vec{\nabla}\cdot \vec{v}$ and using $\Big(\partial (\int \rho \mathcal{J}_L d^3x_0) / \partial t\Big)_L=0$ for an arbitrary density $\rho$ that is non-zero only in a finite spatial region.  The implication is that
\begin{equation}
\left( \frac{\partial \mathcal{J}_L}{\partial t}\right)_L=\mathcal{J}_L\vec{\nabla}\cdot\vec{v}. \label{Lag-Jacobian-ev}
\end{equation}
When $\vec{\nabla}\cdot\vec{v}\neq0$, Equation (\ref{fluid energy}) for the energy evolution can be written for a monatomic ideal gas as
\begin{equation}
\left(\frac{\partial \mathcal{J}_L^{2/3}T}{\partial t}\right)_L = \frac{\mathcal{J}_L^{2/3}T}{m_p}\left(\frac{\partial s_p}{\partial t}\right)_L,
\end{equation}
but no use of this equation will be made in this paper.

 In almost all natural flows, the largest singular value $\Lambda_u$  increases exponentially in time, the smallest $\Lambda_s$ decreases exponentially, and the middle singular value $\Lambda_m$ is slowly varying.   See Figure \ref{fig:exp-fig}b for an illustration of how contours shrink in one direction and stretch in another.  Table I in Reference \cite{Boozer:null-X} gives the singular values for a map that is limited in both the spatial and the angular distances that trajectories can cover, which makes the map particularly interesting for the representation of the velocity of the flowing perfect conductor of Figure \ref{fig:exp-fig}c when modeling the solar corona.  The effect on reconnection of a flow that has similar spatial limitations is studied in the Boozer-Elder paper \cite{Boozer-Elder}.

\subsubsection{Temperature equilibration in Lagrangian coordinates}

The advective-diffusion equation, Equation (\ref{T-ev}) with $2\vec{q}_d/3= - \vec{\nabla}\cdot( D_t\vec{\nabla}T)$, can be solved in Lagrangian coordinates \cite{Tang-Boozer:1999}.  The left hand side is just $(\partial T/\partial t)_L$.  The right hand side can be written using the theory of general coordinates, Appendix \cite{Boozer:NA2015}, as
\begin{eqnarray}
\vec{\nabla}\cdot\vec{q}_d&=& \frac{1}{\mathcal{J}_L } \sum_\alpha \frac{\partial}{\partial x_0^\alpha}\Big(\mathcal{J}_L \vec{\nabla}x_0^\alpha\cdot\vec{q}_d \Big).\\
\vec{\nabla}x_0^\alpha\cdot \frac{2}{3}\vec{q}_d &=& -D_T\vec{\nabla}x_0^\alpha\cdot \vec{\nabla}T  \\
 &=&- D_T \sum_\beta \vec{\nabla}x_0^\alpha \cdot \vec{\nabla}x_0^\beta \frac{\partial T}{\partial x_0^\beta}.
 \end{eqnarray}
Equation (\ref{grad T}) for $\vec{\nabla}x_0^\beta$ and its adjoint for $\vec{\nabla}x_0^\alpha$ imply the inverse metric tensor
\begin{eqnarray}
g^{\alpha \beta}&\equiv& \vec{\nabla}x_0^\alpha \cdot \vec{\nabla}x_0^\beta\\
&=& (\vec{\nabla}_0 x_0^\alpha)^\dag \cdot \left(\tensor{J}^{-1}\right) \cdot  \left(\tensor{J}^{-1}\right)^\dag\cdot\vec{\nabla}_0 x_0^\beta \\
&=& \frac{ (\hat{u}^\dag\cdot\vec{\nabla}_0 x_0^\alpha) (\hat{u}^\dag\cdot\vec{\nabla}_0 x_0^\beta) }{\Lambda_u^2} \nonumber\\
&&+  \frac{ (\hat{m}^\dag\cdot\vec{\nabla}_0 x_0^\alpha) (\hat{m}^\dag\cdot\vec{\nabla}_0 x_0^\beta) }{\Lambda_m^2} \nonumber\\
&& +  \frac{ (\hat{s}^\dag\cdot\vec{\nabla}_0 x_0^\alpha) (\hat{s}^\dag\cdot\vec{\nabla}_0 x_0^\beta) }{\Lambda_s^2}\\
&\rightarrow& \frac{ (\hat{s}^\dag\cdot\vec{\nabla}_0 x_0^\alpha) (\hat{s}^\dag\cdot\vec{\nabla}_0 x_0^\beta) }{\Lambda_s^2},
\end{eqnarray}
when $\Lambda_u>>\Lambda_m>>\Lambda_s$.  In this limit
\begin{eqnarray}
&& \int \Big(\frac{\partial T^2}{\partial t} \Big)_L \mathcal{J}_L d^3x_0 = \nonumber\\
&&= - \int 2\frac{\Lambda_u\Lambda_m}{\Lambda_s}D_T (\hat{s}^\dag\cdot\vec{\nabla}_0 T)^2  d^3x_0. 
\end{eqnarray}
The rate of temperature relaxation is enhanced by the factor $\Lambda_u\Lambda_m / \Lambda_s\approx e^{2\gamma_{ev} t/\tau_{ev}}$.


\section{Magnetic reconnection \label{sec:reconnection} }

The magnetic Reynolds number, $R_m$ is the ratio of the advective to the diffusive effects in the evolution of a magnetic field.  Values of $R_m$ are extremely large in problems of practical interest.  In their 2016 review of magnetic reconnection Zweibel and Yamada \cite{Zweibel:review} stated that ``\emph{In moderately large laboratory plasmas $R_m$ is typically of order $10^4 - 10^8$, in the Sun $R_m\sim 10^8-10^{14}$, while in the interstellar medium of galaxies $R_m \sim 10^{15} -10^{21}$.}"  Although the non-ideal effects, $1/R_m$ are extremely small, the non-ideal effects fundamentally change the nature of the solution---only with their retention can the magnetic field lines change their connections.  What appears truly remarkable is that a term as small as $1/R_m$ can cause reconnection on a time scale $\tau_r$ that differs by approximately one order of magnitude from the advective, or ideal-evolution, time scale $\tau_{ev}$.  Lagrangian coordinates allow one to show that $\tau_r/\tau_{ev} \sim \ln(R_m)$; even when $R_m=10^{21}$ its natural logarithm is relatively small, $\ln(10^{21})\approx 48.3$.  That magnetic reconnection will occur on this time scale has the same certainty as that a radiator can warm a room in of order ten minutes rather than in a couple of weeks.

When a magnetic field evolves from a state in which reconnection is negligible on the time scale of the evolution, two distinct time scales are important.  A time $\tau_r$ is required before the rate of reconnection competes with the rate of evolution.   For shorter times, the magnetic evolution is essentially ideal with the magnetic field lines having trajectories of increasing spatial complexity.  

Once reconnection competes with evolution, the severing and reconnection of magnetic field lines produces forces, which relax on an Alfv\'enic time scale, $\tau_A=L/V_A$.  For example when magnetic field lines carrying distinct parallel currents connect \cite{Boozer:j-||}, a large gradient in $j_{||}/B$ arises, which gives a large Lorentz force, $\vec{f}_L\equiv\vec{j}\times\vec{B}$;
\begin{eqnarray}
\vec{B}\cdot\vec{\nabla}\left(\frac{j_{||}}{B}\right)&=&\vec{B}\cdot\vec{\nabla}\times\left(\frac{\vec{f}_L}{B^2}\right)
\end{eqnarray}
follows from $\vec{\nabla}\cdot\vec{j}=0$.  Alfv\'en waves propagating across the magnetic field lines produce an ideal evolution of the magnetic lines, which in general increases the exponential separation of magnetic field lines that come in close proximity to each other, which produces additional reconnection.  When either reconnection occurs on a significant scale or the magnetic field undergoes an ideal instability, large scale reconnection can follow on an Alfv\'enic time scale.


\subsection{Ideal magnetic energy evolution \label{Sec:B-energy}}

As in the case of an ideal gas, the energy equation for a magnetic field places a constraint on the flow velocity to avoid strong forces.  The derivation of the ideal energy evolution equation is simplified by starting with the magnetic Poynting's theorem in which $\vec{\nabla}\times\vec{B}=\mu_0\vec{j}$,
 \begin{eqnarray}
\frac{\partial}{\partial t}\left(\frac{B^2}{2\mu_0}\right) +\vec{\nabla}\cdot\left(\frac{\vec{E}\times\vec{B}}{\mu_0}\right) =- \vec{j}\cdot\vec{E}. \hspace{0.1in} \label{Poynting}
\end{eqnarray}

In an ideal evolution, $\vec{E}+\vec{u}_\bot\times\vec{B}=-\vec{\nabla}\Phi$. The potential $\Phi$ cancels between the two sides of Equation (\ref{Poynting}), and the derivation is simplified by letting $\vec{E}+\vec{u}_\bot\times\vec{B}=0$;  then  $\vec{E}\times \vec{B} = -(\vec{u}_\bot\times\vec{B})\times\vec{B}=B^2\vec{u}_\bot$ and 
\begin{eqnarray}
\vec{j}\cdot\vec{E}&=&\vec{u}_\bot\cdot(\vec{j}\times\vec{B})\\
&=&\vec{u}_\bot\cdot\vec{f}_L, \mbox{   where   }  \\
\vec{f}_L&=&\vec{j}\times\vec{B}
\end{eqnarray}
is the Lorentz force, the force per unit volume a magnetic field exerts on any material carrying a current of density $\vec{j}$. 

The evolution of the magnetic energy density can be placed in a form analogous to Equation (\ref{fluid energy}) for the thermal energy density; the energy conserving terms on the left-hand side of the equation and the non-energy conserving terms on the right-hand side:
\begin{eqnarray}
&&\frac{\partial}{\partial t}\left(\frac{B^2}{2\mu_0}\right) +\vec{\nabla}\cdot\left(\frac{B^2}{\mu_0} \vec{u}_\bot\right)=-\vec{u}_\bot\cdot\vec{f}_L \label{mag-E-tranport}\\
&&\frac{\partial}{\partial t}\left(\frac{B^2}{2\mu_0}\right) +\vec{\nabla}\cdot\left(\frac{B^2}{2\mu_0} \vec{u}_\bot\right)=\nonumber\\
&& \hspace{0.5in} -\Big\{\vec{\nabla}\cdot\left(\frac{B^2}{2\mu_0}\vec{u}_\bot \right) + \vec{u}_\bot\cdot \vec{f}_L \Big\}.
\end{eqnarray}
Equation (\ref{mag-E-tranport}) has the same property as equation for the evolution of the energy density for a gas, Equation (\ref{gas energy}); the energy flux is the energy density plus the pressure times the velocity and not just the energy density times the velocity. 

The power per unit volume associated with the Lorentz force, $\vec{u}_\bot\cdot \vec{f}_L$ does not explicily enter the equation for magnetic energy conservation. Using a vector identity and then Ampere's law, $\vec{\nabla}\times\vec{B}=\mu_0\vec{j}$,
\begin{eqnarray}
&& \vec{\nabla}B^2= -2(\vec{\nabla}\times\vec{B})\times\vec{B}  + 2 \vec{B}\cdot\vec{\nabla}\vec{B}, \mbox{    so   }\\
&& \vec{\nabla}\left(\frac{B^2}{2\mu_0}\right) = \left(\frac{B^2}{2\mu_0}\right)\hat{b} +  2\left(\frac{B^2}{2\mu_0}\right)\vec{\kappa} - \vec{f}_L;\\
&& \vec{\nabla}\left(\frac{B^2}{2\mu_0}\vec{u}_\bot\right) = \left(\frac{B^2}{2\mu_0}\right)\left(\vec{\nabla}\cdot\vec{u}_\bot +2 \vec{u}_\bot\cdot\vec{\kappa}\right)\nonumber\\
&& \hspace{1.0in} - \vec{u}_\bot\cdot \vec{f}_L. 
\end{eqnarray}
The curvature of the magnetic field lines is $\vec{\kappa}=\hat{b}\cdot\vec{\nabla}\hat{b}$ where $\hat{b}\equiv \vec{B}/B$ is the unit vector along $\vec{B}$.  The evolution of the magnetic energy is then
\begin{eqnarray}
&&\frac{\partial}{\partial t}\left(\frac{B^2}{2\mu_0}\right) +\vec{\nabla}\cdot\left(\frac{B^2}{2\mu_0} \vec{u}_\bot\right) \nonumber\\
&& \hspace{0.3in} =-\left(\frac{B^2}{2\mu_0}\right)\big(\vec{\nabla}\cdot\vec{u}_\bot+2\vec{u}_\bot\cdot\vec{\kappa}\big). \label{Eq:B-energy}
\end{eqnarray}
Equation (\ref{Eq:B-energy}) has the same form as Equation (\ref{fluid energy}) for the thermal energy density but with $\vec{\nabla}\cdot\vec{v}$ replaced by $\vec{\nabla}\cdot\vec{u}_\bot+2\vec{u}_\bot\cdot\vec{\kappa}$.

In the thermal or in the magnetic case, a violation of the constraint $\vec{\nabla}\cdot\vec{v}=0$ or $\vec{\nabla}\cdot\vec{u}_\bot+2\vec{u}_\bot\cdot\vec{\kappa}=0$ implies the presence of a force that transfers either thermal or magnetic energy.  Stated the other way, a thermal system can be modified with no energy transfer by a velocity that is arbitrary other than $\vec{\nabla}\cdot\vec{v}=0$ and the magnetic field lines can be modified with no energy transfer by moving the lines with a velocity $\vec{u}_\bot$ that is arbitrary other than  $\vec{\nabla}\cdot\vec{u}_\bot+2\vec{u}_\bot\cdot\vec{\kappa}=0$.  Both constraints are naturally obeyed for systems evolving slowly compared to the transit time, for sound waves for the thermal system or Alfv\'en waves for the magnetic system.

The important question for the thermal system is whether temperature gradients can be relaxed exponentially faster by increasing the temperature gradient by a flow satisfying $\vec{\nabla}\cdot\vec{v}=0$.  The answer is positive in systems with at least two spatial dimensions as discussed in Section \ref{sec:T-equil}, but negative in systems with only one spatial dimension, for then $\vec{\nabla}\cdot\vec{v}=0$ implies a constant speed.

The important question for magnetic reconnection is whether the speed at which magnetic field lines change their topology can be exponentially enhanced by  the distortions of tubes of magnetic flux by a magnetic field line flow satisfying $\vec{\nabla}\cdot\vec{u}_\bot+2\vec{u}_\bot\cdot\vec{\kappa}=0$.  The answer is positive in systems with at least three spatial dimensions: one must be along the magnetic field and two must be perpendicular to satisfy the constraint on $\vec{u}_\bot$.

There is an important difference between the thermal and the magnetic system.  An ideal thermal evolution with a chaotic flow causes the magnitude of the temperature gradient to increase exponentially in time.   But, an ideal magnetic evolution with a chaotic flow of the magnetic field lines does not cause the parallel current $j_{||}$ to increase exponentially, Equation (\ref{K}).  See Boozer and Elder \cite{Boozer-Elder} for a solved reconnection example.  It is the distortion of the tubes of magnetic flux that causes an exponential enhancement in the rate of reconnection not an exponential enhancement in $E_{||}=\eta j_{||}$.  Although magnetic reconnection can release a significant fraction of the energy in the large-scale magnetic field into Alfv\'en waves, an exponentially small fraction of the energy is directly dissipated \cite{Boozer:part.acc}.

Note that when $\vec{\nabla}\cdot\vec{u}_\bot+2\vec{u}_\bot\cdot\vec{\kappa}=0$, Equation (\ref{Eq:B-energy}) for the ideal magnetic-energy evolution and Equation (\ref{Lag-Jacobian-ev}), which says $(\partial \mathcal{J}_L/\partial t)_L = \mathcal{J}_L\vec{\nabla}\cdot\vec{u}_\bot$, imply  
\begin{eqnarray}
&&\left(\frac{\partial (\mathcal{J}_LB^2)}{\partial t} \right)_L=0 \hspace{0.2in} \mbox{and   } \\
&& \left(\frac{\partial \ln(\mathcal{J}_L)}{\partial t} \right)_L =-2\vec{u}_\bot\cdot\kappa.
\end{eqnarray}



\subsection{ Ideal $\vec{B}$ evolution in Lagrangian coordinates}

A vector identity implies the ideal evolution equation for $\vec{B}$ can be written 
\begin{eqnarray}
\frac{\partial \vec{B} }{\partial t} &=& - \vec{B} \vec{\nabla}\cdot\vec{u} - \vec{u}\cdot\vec{\nabla}\vec{B}+\vec{B}\vec{\nabla}\vec{u}, \mbox{   so  } \hspace{0.2in}\\
\Big(\frac{\partial \mathcal{J}_L\vec{B} }{\partial t} \Big)_L &=& \mathcal{J}_L\vec{B} \cdot\vec{\nabla}\vec{u},
\end{eqnarray}
using the definition of a Lagrangian derivative and Equation (\ref{Lag-Jacobian-ev}) for $(\partial \mathcal{J}_L/\partial t)_L$.  
\begin{eqnarray}
\vec{u}&=& \Big(\frac{\partial \vec{x} }{\partial t} \Big)_L  \mbox{   so  } \\
\frac{\partial \vec{u}}{\partial \vec{x}_0} &=& \Big(\frac{\partial \tensor{J} }{\partial t} \Big)_L \\
\vec{B}\cdot\vec{\nabla}\vec{u}&=& \Big(\frac{\partial \tensor{J} }{\partial t} \Big)_L\cdot\tensor{J}^{-1}\cdot\vec{B}, \mbox{   so   } \\
\vec{B}(\vec{x},t) &=& \frac{\tensor{J}}{\mathcal{J}}\cdot \vec{B}_0. \label{Lag-B-ev}
\end{eqnarray}
$\vec{B}_0$ is the magnetic field at $t=0$.  The derivation and the history of this form for the  field was reviewed in 1966 by Stern \cite{Stern:1966} and mentioned in \cite{Zweibel:review}.

The ideal magnetic evolution in Lagrangian coordinates, Equation (\ref{Lag-B-ev}), is valid even when the constraint $\vec{\nabla}\cdot\vec{u}_\bot+2\vec{u}_\bot\cdot\vec{\kappa}$ is non-zero.  This is unlike the analogous result for an ideal thermal evolution, which requires $\vec{\nabla}\cdot\vec{v}=0$.

Equation (\ref{Lag-B-ev}) implies that
\begin{eqnarray}
B^2 = \left( \frac{\hat{u}^\dag\cdot\vec{B}_0}{\Lambda_m\Lambda_s}\right)^2 +  \left( \frac{\hat{m}^\dag\cdot\vec{B}_0}{\Lambda_u\Lambda_s}\right)^2+ \left( \frac{\hat{s}^\dag\cdot\vec{B}_0}{\Lambda_u\Lambda_m}\right)^2.\hspace{0.06in}
\end{eqnarray}
The term in $B^2 $ proportional to $(\hat{u}^\dag\cdot\vec{B}_0)^2$ goes to infinity exponentially in time.   The term  proportional to $(\hat{s}^\dag\cdot\vec{B}_0)^2$ goes to zero exponentially.   A bounded magnetic field strength is only possible for a time long compared to $\tau_{ev}$ when the effective magnetic field points in the $\hat{M}$ direction, 
\begin{equation}
\vec{B}(\vec{x},t) \rightarrow \frac{\hat{m}^\dag\cdot\vec{B}_0}{\Lambda_u\Lambda_s}\hat{M}.
\end{equation}
The unit vector $\hat{M}$ is also the unit vector along the magnetic field $\hat{b}$.

  

\subsection{Clebsch potentials and the evolution of $\vec{B}$ \label{sec:B-ev} }

In 1958 Newcomb \cite{Newcomb} studied the behavior of magnetic field lines given by Equation (\ref{ideal ev B}) of an ideal evolution. He showed that the magnetic field lines move with a velocity $\vec{u}$ and cannot break.  His results can be generalized to study the behavior of the magnetic field lines in an arbitrary evolution. 

A divergence-free field, such as the magnetic field, can be written in the Clebsch potentials $\alpha(\vec{x},t)$ and $\beta$, 
\begin{eqnarray}
\vec{B} &=&\vec{\nabla}\alpha\times\vec{\nabla}\beta. \label{Clebsch}
\end{eqnarray}
The history of this form was reviewed by Stern \cite{Stern:1970}, who calls $\alpha$ and $\beta$ Euler potentials, but the name Clebsch potentials is more common among plasma physicists.  The time derivative of $\vec{B}$ is given by
\begin{eqnarray}
&& \frac{\partial \vec{B}}{\partial t} =\vec{\nabla}\times\left(\vec{u}\times\vec{B} +\frac{\partial g}{\partial\ell} \vec{\nabla}\ell \right); \\
&& \frac{\partial \alpha}{\partial t}+\vec{u}\cdot\vec{\nabla}\alpha = \frac{\partial g}{\partial\beta}; \\
&& \frac{\partial \beta}{\partial t}+\vec{u}\cdot\vec{\nabla}\beta =- \frac{\partial g}{\partial\alpha},
\end{eqnarray}
where $\ell$ is the distance along a magnetic field line.  The function $g$ in principle depends on $(\alpha,\beta,\ell,t)$, but the evolution is ideal only when $\partial g/\partial\ell=0$.  A $g$ that has no $\ell$ dependence is designated  as ideal, $g_{I}(\alpha,\beta,t)$.

The arbitrary function $g_{I}(\alpha,\beta,t)$ is equivalent to an arbitrary velocity $\vec{u}_a$ that can be added to $\vec{u}_\bot$ in an ideal evolution,
\begin{equation} 
\vec{u}_a\equiv\frac{\vec{B}\times\vec{\nabla}g_{I}(\alpha,\beta,t)}{B^2}. \label{u_a}
\end{equation} 
Then, $\vec{u}_a\cdot\vec{\nabla}\alpha = -\partial g_I/\partial \beta$, $\vec{u}_a\cdot\vec{\nabla}\beta = \partial g_I/\partial \alpha$, and $\vec{\nabla}\cdot(B^2\vec{u}_a/\mu_0)=- \vec{u}_a\cdot\vec{f}_L$.    
  

\subsection{The current density in an ideal evolution }

Both analytic and numerical models of reconnection commonly assume the initial state of the field contains a current sheet with a current density $j\propto R_m$, which gives rapid reconnection, but how a magnetic field could evolve into such an extreme state is not explained.  

Magnetic reconnection is so prevalent that even an initially curl-free field, $\vec{B}_0 = \vec{\nabla}_0\phi_0$, must be able to reach a state through an ideal evolution in which reconnection competes with the ideal evolution.  As will be shown, this naturally occurs as $\Lambda_u\propto e^{\gamma_{ev}t/\tau_{ev}}\rightarrow\infty$.  For a specific reconnection example, see Boozer and Elder \cite{Boozer-Elder}.

A magnetic field that has undergone an ideal evolution from an initially curl-free state obeys
\begin{equation}
\vec{B} = \frac{\Lambda_u \hat{u}^\dag\cdot \vec{\nabla}_0\phi_0}{\mathcal{J}_L} \hat{U} + \frac{\Lambda_m \hat{m}^\dag\cdot \vec{\nabla}_0\phi_0}{\mathcal{J}_L} \hat{M} + \frac{\Lambda_s \hat{s}^\dag\cdot \vec{\nabla}_0\phi_0}{\mathcal{J}_L} \hat{S}.
\end{equation}
The finiteness $\vec{B}$ as $\Lambda_u\rightarrow\infty$ implies $\hat{u}^\dag\cdot \vec{\nabla}_0\phi=0$ and that $\vec{B}=B\hat{M}$, so as $\Lambda_u\rightarrow\infty$ and $\Lambda_s\rightarrow0$,
\begin{eqnarray}
\vec{B} &=& B \hat{M} + \frac{\Lambda_s \hat{s}^\dag\cdot \vec{\nabla}_0\phi_0}{\mathcal{J}_L} \hat{S};\\
B &=& \frac{\Lambda_m \hat{m}^\dag\cdot \vec{\nabla}_0\phi_0}{\mathcal{J}_L} \label{Mag-B}.
\end{eqnarray}

The potential $\phi$ can be defined for non-zero time by $(\partial \phi/\partial t)_L=0$, then
Equation  (\ref{grad T}) for gradients and Equation (\ref{Mag-B}) imply
\begin{eqnarray}
\vec{\nabla}\phi &=& \frac{ \hat{m}^\dag\cdot \vec{\nabla}_0\phi_0}{\Lambda_m} \hat{M} + \frac{ \hat{s}^\dag\cdot \vec{\nabla}_0\phi_0}{\Lambda_s} \hat{S} \\
&=&\frac{\mathcal{J}_L B}{\Lambda_m^2} \hat{M} + \frac{ \hat{s}^\dag\cdot \vec{\nabla}_0\phi_0}{\Lambda_s} \hat{S}. \label{phi-M-S}
\end{eqnarray}

The freedom within an ideal evolution, Equation (\ref{u_a}), can be used to ensure the Clebsch coordinate $\beta$ satisfies $\hat{u}^\dag\cdot\vec{\nabla}_0\beta_0=0$.  Since $\vec{B}\cdot \vec{\nabla}\beta=0$, which implies $\hat{M}\cdot \vec{\nabla}\beta=0$, Equation  (\ref{grad T}) for gradients of functions carried by the flow implies that as $\Lambda_u\rightarrow\infty$ the gradient of $\beta$ is
\begin{equation}
\vec{\nabla}\beta = \frac{ \hat{s}^\dag\cdot \vec{\nabla}_0\beta_0}{\Lambda_s} \hat{S}. \label{grad beta}
\end{equation}

Equations (\ref{phi-M-S}) and (\ref{grad beta}) imply that the magnetic field, $\vec{B}=B\hat{M}$, has the covariant form
\begin{eqnarray}
\vec{B} &=& \frac{\Lambda_m^2}{\mathcal{J}_L} \left(\vec{\nabla}\phi - \frac{ \hat{s}^\dag\cdot \vec{\nabla}_0\phi_0}{\hat{s}^\dag\cdot \vec{\nabla}_0\beta_0}\vec{\nabla}\beta \right) \\
&=& B_\phi \vec{\nabla}\phi +B_\beta \vec{\nabla}\beta; \\
\vec{B}\cdot\vec{\nabla}\times\vec{B} &=& B_\phi^2\frac{\partial (B_\beta/B_\phi)}{\partial\alpha} (\vec{\nabla}\alpha\times\vec{\nabla}\beta)\cdot\vec{\nabla}\phi. \hspace{0.2in}
\end{eqnarray}
The triple product $(\vec{\nabla}\alpha\times\vec{\nabla}\beta)\cdot\vec{\nabla}\phi = B\hat{M}\cdot\vec{\nabla}\phi = B^2/B_\phi$, which implies
\begin{eqnarray}
K&\equiv& \frac{\mu_0j_{||}}{B} = \frac{ \vec{B} \cdot \vec{\nabla}\times \vec{B} }{B^2} \\
&=&B_\phi \frac{\partial (B_\beta/B_\phi)}{\partial\alpha} \nonumber\\
&=& - \frac{\Lambda_m^2}{\mathcal{J}_L}  \frac{\partial }{\partial\alpha}\left( \frac{ \hat{s}^\dag\cdot \vec{\nabla}_0\phi}{\hat{s}^\dag\cdot \vec{\nabla}_0\beta}\right). \label{K}
\end{eqnarray}
The absence of a strong Lorentz force requires $\partial K/\partial\phi=0$.   Equation (\ref{K}) for $K$ is corrected from the expression given in \cite{Boozer:ideal-ev}.

Equation (\ref{K}) does not require an exponential growth in the force-free or parallel current density.  Equation  (\ref{grad T}) for gradients implies
\begin{eqnarray}
\vec{\nabla}K &=& \frac{ \hat{u}^\dag\cdot \vec{\nabla}_0K}{\Lambda_u}\hat{U} +\frac{ \hat{m}^\dag\cdot \vec{\nabla}_0K}{\Lambda_m}\hat{M}+\frac{ \hat{s}^\dag\cdot \vec{\nabla}_0K}{\Lambda_s}\hat{S}. \nonumber\\
\end{eqnarray}
The parallel current distribution $K$ lies in sheets that are very extended in the direction $\hat{U}$ in which streamlines of $\vec{u}_\bot$ exponentially separate from each other and very narrow in the direction $\hat{U}$ in which streamlines of $\vec{u}_\bot$ exponentially approach each other.
\begin{eqnarray}
&&\vec{\nabla}\alpha = \frac {\hat{U}}{\Lambda_u B_\beta}  + \frac{ \hat{s}^\dag\cdot \vec{\nabla}_0\alpha }{\Lambda_s} \hat{S},\\
&&\vec{\nabla}K = \frac{\partial K}{\partial\alpha} \frac {\hat{U}}{\Lambda_u B_\beta} + \frac{ \hat{s}^\dag\cdot \vec{\nabla}_0 K }{\Lambda_s} \hat{S}.
\end{eqnarray}

A covariant representation of $\vec{B}$ in $\alpha,\beta,\phi$ coordinates can be obtained from a relation in the theory of general coordinates,
\begin{eqnarray}
\left(\frac{\partial \vec{x}}{\partial\phi}\right)_{\alpha \beta} &=& \frac{\vec{\nabla}\alpha\times\vec{\nabla}\beta}{(\vec{\nabla}\alpha\times\vec{\nabla}\beta)\cdot\vec{\nabla}\phi} \\
&=& \frac{B_\phi}{B^2}\vec{B},  \hspace{0.2in}\mbox{so   } \\
\vec{B} &=& \frac{\mathcal{J}_L}{\Lambda_m^2} B^2 \frac{\partial \vec{x}}{\partial\phi}.
\end{eqnarray}

\vspace{0.1in}
\subsection{Required current density for reconnection}

Contrary to the conventional view, the current density need not be large to obtain the non-ideal electric field $\mathcal{E}_{ni}$ required for a rapid reconnection.  The current-density requirement comes not from the magnitude of $\mathcal{E}_{ni}$ but from the requirement for sufficient  distortion in the magnetic flux tubes.  As discussed in \cite{Boozer:null-X}, the current density need not be large to cause magnetic field lines to exponentiate apart; it needs to increase only linearly in the number of exponentiations \cite{Boozer:B-line.sep}.   This is explicitly shown for the reconnection example of Boozer and Elder \cite{Boozer-Elder}.

An argument similar to the one that led to Equation (\ref{delta-ev}) implies the separation $\vec{\Delta}$ between magnetic field lines changes as $d\vec{\Delta}/d\ell=\vec{\Delta}\cdot\vec{\nabla}\hat{b}$.  A parallel current produces a $\Big| \vec{\nabla}\hat{b} \Big|\approx \mu_0 j_{||}/B$.  The distance required for an e-fold in separation is only a few times longer than $1/K$, where $K\equiv\mu_0 j_{||}/B$.  To have $R_m$ e-folds within a distance $L$ along the field lines requires $KL > \ln(R_m)$, but only a few times greater.  This is in contrast to the current density required in the traditional assumption that $\mathcal{E}_{ni}$ must compete with $\Big| \vec{u}\times\vec{B} \Big|$, which implies $KL\approx R_m$ in places where reconnection is occurring.


\subsection{Non-ideal magnetic-field evolution}

Two types of effects limit the ideal evolution and produce magnetic reconnection.  The most universal effect is electron inertia, which makes magnetic field lines that approach each other closer than $c/\omega_{pe}$ anywhere on their trajectories indistinguishable in an evolution, Appendix C of  \cite{Boozer:null-X}.  The distance $c/\omega_{pe}$ acts in a way that is analogous to the mesh size limiting the resolution in a numerical calculation.  Indeed, the finite size of the mesh in a numerical simulation produces reconnection even when explicit non-ideal effects are ignored \cite{Pariat-Antiochos}.  The second and more commonly discussed effect is the plasma resistivity, which causes a diffusion of the magnetic file lines with a diffusion coefficient $\eta/\mu_0$.  Both effects are small.  In the solar corona a typical electron density is $10^{14}/$m$^3$, for which $c/\omega_{pe}\approx 0.5$~m, while typical distance scales are of order $10^8$~m.  Resistive effects measured by $1/R_m$ are even smaller in the corona.

The natural mathematical description of two non-ideal effects is distinct and can be addressed by defining the effective  magnetic field $\vec{\mathcal{B}}$, which evolves as if $c/\omega_{pe}$ were zero, from which the actual magnetic field $\vec{B}$ can be obtained.

\subsubsection{Representation of generalized Ohm's law \label{sec:Ohm's law} }

As discussed in the Introduction and in \cite{Boozer:part.acc}, the velocity of the magnetic field lines $\vec{u}_\bot$ and the velocity of the plasma $\vec{v}$ in which the magnetic field is embedded are distinct.  Unlike in these discussions, the electron inertial term in Ohm's law, which is proportional to $\partial\vec{j}/\partial t$, will be treated separately since it has special mathematical properties.

Any generalization of Ohm's law that does not contain integrals over space or time can be written in a plasma moving with a velocity $\vec{v}$ as 
\begin{eqnarray}
&& \vec{E} + \vec{v}\times\vec{B} = \left(\frac{c}{\omega_{pe}}\right)^2 \mu_0\frac{\partial \vec{j}}{\partial t} +\vec{\mathcal{R}}. \label{E1} 
\end{eqnarray}
The $\partial\vec{j}/\partial t$ term is due to the inertia of the lightest current-carrying particle, which is the electron; $\omega_{pe}\equiv \sqrt{ne^2/\epsilon_0 m_e}$, where $n$ is the number density of electrons of mass $m_e$.  

The magnetic evolution $\partial \vec{B}/\partial t=-\vec{\nabla}\times\vec{E}$ is simplified by defining an effective magnetic field
\begin{eqnarray}
&& \vec{\mathcal{B}} \equiv \vec{B} + \vec{\nabla}\times \left( \left(\frac{c}{\omega_{pe}}\right)^2\vec{\nabla}\times \vec{B} \right) \label{eff.B}
\end{eqnarray}
using $\vec{\nabla}\times\vec{B}=\mu_0\vec{j}$.  Using the effective magnetic field $\vec{\mathcal{B}}$, Equation (\ref{E1}) can be rewritten as
\begin{eqnarray}
&& \vec{E} + \vec{u}\times\vec{\mathcal{B}} = \left(\frac{c}{\omega_{pe}}\right)^2 \mu_0\frac{\partial \vec{j}}{\partial t} -\vec{\nabla}\Phi+\mathcal{E}_{ni}\vec{\nabla}\ell.  \label{E2} \hspace{0.3in}
\end{eqnarray}
 The $\vec{\mathcal{B}}\times$ components of Equation (\ref{E1})  are balanced by $\vec{u}\times\vec{\mathcal{B}}$, which defines the velocity $\vec{u}$.  The $\vec{\mathcal{B}}\cdot$ component can be partially balanced by $\vec{\mathcal{B}}\cdot\vec{\nabla}\Phi$, but $\Phi$ must be a well-behaved, single-valued potential.  The non-ideal electric field $\mathcal{E}_{ni}\vec{\nabla}\ell$, where $\ell$ is the distance along an effective magnetic field line, is introduced to make this possible.  

$\mathcal{E}_{ni}$ is constant along the effective magnetic field lines and is chosen to obtain the correct conditions at boundaries and null points \cite{Boozer:null-X} or for the loop voltage in a torus; 
\begin{equation}
\mathcal{E}_{ni} \equiv \frac{\int \vec{E}\cdot\frac{\vec{B}}{B}d\ell}{\int d\ell}.
\end{equation}
Both integrals are calculated using the same limits of integration.  The integration limits can be (1) $\ell\rightarrow \pm\infty$ as on the irrational magnetic surfaces of a toroidal plasma, (2) a wall on which $\Phi$ has a specified value, such as $\Phi=0$ on a perfectly conducting grounded wall, or (3) the potential $\Phi_0$ on the infinitesimal sphere surrounding a null.   That potential is determined by the condition that no net current enter or leave the null.

\subsubsection{Evolution of the non-ideal magnetic field}

The evolution equation for the effective magnetic field is
\begin{eqnarray}
&& \frac{\partial \vec{\mathcal{B}} }{\partial t} =\vec{\nabla}\left(\vec{u}\times\vec{\mathcal{B}} -\mathcal{E}_{ni}\vec{\nabla}\ell \right).  \label{eff B ev}
\end{eqnarray}
When $\vec{\mathcal{B}}(\vec{x},t)$ is known, Equation (\ref{eff.B}) can be solved for the true magnetic field $\vec{B}(\vec{x},t)$.   As shown in Appendix C of  \cite{Boozer:null-X}, the $\vec{B}$ is $\vec{\mathcal{B}}$ dispersed by a distance $c/\omega_{pe}$ across the lines.


While effects due to $\mathcal{E}_{ni}$ are small, the effective magnetic field can be taken to be an ideally evolving field plus a non-ideal field, $\vec{\mathcal{B}} = \vec{\mathcal{B}}_I +\vec{\mathcal{B}}_{ni}$.  Retaining only the first order deviation from an ideal evolution,
\begin{eqnarray}
\vec{\mathcal{B}}_I &=& \vec{\nabla}\alpha_I \times  \vec{\nabla}\beta_I ; \\
\alpha &=& \alpha_I - \frac{\partial (\mathcal{A}_{ni}\ell)}{\partial\beta_I}; \\
\beta &=& \beta_I + \frac{\partial (\mathcal{A}_{ni}\ell) }{\partial\alpha_I};\\
\mathcal{A}_{ni} &\equiv& - \int_0^t \mathcal{E}_{ni}dt; \\
\vec{\mathcal{B}}_{ni} &=& \vec{\nabla}\mathcal{A}_{ni}\times\vec{\nabla}\ell. \hspace{0.2in} \label{B_ni}
\end{eqnarray}
$\mathcal{A}_{ni}(\alpha_I,\beta_I,t)\vec{\nabla}\ell$ is the non-ideal part of the vector potential.


  Equation (\ref{grad T}) for the gradient of a function has the asymptotic form $\vec{\nabla}f \rightarrow \left(\hat{s}\cdot \vec{\nabla}_0f/\Lambda_s\right)\hat{S}$, and becomes exponentially large in the direction in which streamlines of $\vec{u}_\bot$ approach each other; $\Lambda_s$ goes to zero exponentially.  Consequently $\vec{\nabla}\mathcal{A}_{ni} \rightarrow \hat{S}(\hat{s}^\dag\cdot\vec{\nabla}_0\mathcal{A}_{ni})/\Lambda_s$. The magnetic field lines become oriented in the $\hat{M}$ direction, and $\vec{B}\cdot\vec{\nabla}\ell=B$ implies $\vec{\nabla}\ell $ equals $\hat{M}$ plus terms other terms; the term in the $\hat{S}$ direction can be large.  Equation (\ref{B_ni}) then shows that the non-ideal part of the magnetic field grows exponentially in time \cite{Boozer:ideal-ev},
\begin{eqnarray}
\vec{\mathcal{B}}_{ni}&\rightarrow&\frac{\hat{S}\times\hat{M}}{\Lambda_s} \hat{s}^\dag\cdot\vec{\nabla}_0\mathcal{A}_{ni}(\alpha_I,\beta_I,t) \nonumber\\
&=& -\frac{\hat{U}}{\Lambda_s} \hat{s}^\dag \cdot\vec{\nabla}_0\mathcal{A}_{ni}(\alpha_I,\beta_I,t)
\end{eqnarray}
as $\Lambda_s$ approaches zero exponentially.  On a time scale, $\sim\tau_{ev}\ln R_m$ the effective magnetic field will enter a state of fast magnetic reconnection.  This is demonstrated by the reconnection example of Boozer and Elder \cite{Boozer-Elder}.


\section{Discussion  \label{sec:history}}

The enhancement of mixing by stirring is such a part of everyday life that its physical reality cannot be denied.  Stirring produces a topology-conserving motion of fluid elements that have certain composition or temperature.  Mixing implies destroying that topology.   

Remarkably, a mathematical explanation for the enhancement of mixing by stirring did not exist until Aref's development \cite{Aref;1984} of Lagrangian methods  in 1984 for studying  the advective or stirring part of the advection-diffusion equation.  The complete solution of the advection-diffusion equation in Lagrangian coordinates was not given until fifteen years later by Tang and Boozer \cite{Tang-Boozer:1999}.  

In advection-diffusion problems with very weak diffusion, the time for mixing is significantly longer than the evolution time defined by the stirring.  Nevertheless, the mixing time is only an order of magnitude longer even when the processes that allow mixing are many orders of magnitude smaller than the stirring.  This is true independent of the initial state or how the system is stirred, with a few exceptions.  These exceptions are stirring motions that obey symmetries, but even these exceptional stirring motions must be performed with extreme care to avoid mixing.

The evolution of magnetic fields in highly conducting plasmas have obvious resemblances to the problem of stirring.  Independent of the initial state, the magnetic field lines move with a topology conserving velocity for an evolution time defined by that velocity.  On a time scale approximately an order of magnitude longer than the stirring time, the topology of the magnetic field lines is destroyed.  This is true even when characteristic amplitude of the topology-destroying terms is ten orders of magnitude smaller than the advective terms.  The same Lagrangian methods used to explain fluid mixing also explain magnetic reconnection with the same degree of certainty.  

The representation of the ideal evolution of magnetic fields in Lagrangian coordinates has long been known---the history of that representation was reviewed eighteen years before Aref's paper, in 1966 by David Stern \cite{Stern:1966}, who ascribed the representation to an 1816 paper by A.-L. Cauchy on vorticity evolution.  Nevertheless, the paradigms that have been developed during more than sixty years of intense study of magnetic reconnection are so disconnected from the explanations that follow naturally using Lagrangian coordinates that these results have been largely ignored rather than used for analyses. This disconnection is probably due to a pervasive assumption that reconnection can be understood in two dimensions even though the problems of interest are in three dimensional space.  For ordinary fluids, stirring leads to mixing to two dimensions but this is not true for magnetic fields. 


James Dungey had insights that were close to results discussed in this paper.  According to the 2016 tribute in Eos \cite{Southwood:2016}, Dungey was struck by the importance of the ``\emph{pattern milk made as it was stirred into the coffee}" while sitting in a Parisian caf\'e, which led to his 1961 paper \cite{Dungey:1961} that laid the foundations of the magnetospheric models.   Much earlier, in 1953, Dungey developed \cite{Dungey:1953} the theory of  both the ideal motion, $\vec{u}$, and the breaking of magnetic field lines due to a loop voltage  $V=\oint \vec{E}\cdot d\vec{\ell}$ with $d\vec{\ell}$ along $\vec{B}$.  Remarkably his focus was on toroidal magnetic surfaces.  Dungey noted the Hall effect is perpendicular to $\vec{B}$, so it has no direct effect on magnetic field line breaking, and that resistive breaking is very slow in astrophysical systems but that the speed could be increased by turbulence. 

The basic problem with the resistive time scale was recognized but not explicitly given by Dungey.  When a region with a continuous symmetry that has a length $L$ and a width $a<<L$ changes the magnetic flux $\psi_{rec}=B_{rec} La$ by a reconnection process, then rate of flux change is $\oint\vec{E}\cdot d\vec{x} \approx \eta j L$.  The time scale for the flux change $\tau_\psi\approx B_{rec} a/\eta j$, but this must compete with the evolution time $\tau_{ev}=a/u$ to be of significance.  When the current flows in a channel of width $\Delta_d$, the current density $j\approx B/\mu_0\Delta$.  The requirement that $\tau_\psi$ be less than or equal to $\tau_{ev}$, is that $\Delta_d \lesssim a/R_m$.

A common picture for obtaining an intense current density is to assume the magnetic field lies in quasi-discrete bundles, which can be called a flux ropes, with an intense magnetic field inside the rope and a negligible field outside.  In astrophysics these bundles are often called flux tubes.  Indeed they are tubes of flux, but there should be no implication that the field is unusually strong in the interior of a magnetic flux tube.  

When two flux ropes collide, an intense current appears at the point of collision and this was shown by Sweet and Parker \cite{Sweet:1958,Parker;1957} to lead to a current density enhanced by a factor of $\sqrt{R_m}$, but this enhancement is too small to compete with evolution.  Most of the reconnection literature since that time has been an effort to identify a mechanism for obtaining a sufficiently intense current density as discussed in reviews \cite{Zweibel:review,Loureiro:2016}.    

Other than the 1988 paper of Schindler, Hesse, and Birn \cite{Schindler:1988}, remarkably little has been written about reconnection needing to complete with evolution to be an important process.  More emphasis has been given to the speed of observed reconnection phenomena, which Parker noted in 1973 were \emph{``universally of the general order of magnitude of $0.1~V_A$,"} \cite{Parker:1973} where $V_A$ is the Alfv\'en speed.  The observed Alfv\'enic rate has little implication on the intrinsic cause for reconnection. When magnetic field lines  break more rapidly than an externally driven evolution, quasi-static force balance is generally lost; inertial forces  provide force balance, which implies an internal evolution rate determined by the speed of Alfv\'en waves.

\begin{figure}
\centerline{ \includegraphics[width=2.5in]{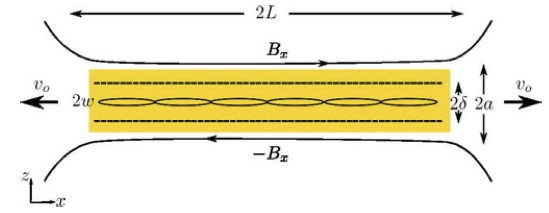}}
\caption{In plasmoid models, oppositely directed fields in the $B_x$ direction are pushed together forming a narrow current sheet.  Tearing instability of the sheet current creates the plasmoids, which are expelled at the Alfv\'en speed from two ends of the current sheet.  The maximum current density in the reconnection or plasmoid region is proportional to $R_m$.  This figure is from Reference \cite{Huang-Comisso}. }
\label{fig:plasmoid}
\end{figure}

The traditional interpretation of the difficulty of achieving the observed speed of reconnection \cite{Zweibel:review,Loureiro:2016} is that plasma must be rapidly removed from the reconnection region to maintain the current density $j\propto R_m$ that is needed in two dimensional models.  The method that dominates the modern literature is Alfv\'enic expulsion of plasmoids \cite{Loureiro:2016} along the thin channel in which reconnection takes place, Figure \ref{fig:plasmoid}, which is from \cite{Huang-Comisso}.  The rapid removal of plasma is not an issue in a fully three-dimensional reconnection but is in two.

In 1994 Longcope and Strauss \cite{Longcope-Strauss:1994} studied the effect of an imposed magnetic field line flow on the formation of strong current in a two-dimensional, time dependent model.   Although two spatial dimensions is sufficient for the exponentially enhanced mixing of fluids, it is not adequate for the enhancement of magnetic reconnection.  The reason is clear using Equation (\ref{Eq:B-energy}), for the evolution of the energy density in the magnetic field during an ideal evolution.  The magnetic energy density increases exponentially in time at a rate $\nu_B=\vec{\nabla}\cdot\vec{u}_\bot+2\vec{u}_\bot\cdot\vec{\kappa}$, where $\vec{\kappa}\equiv\hat{b}\cdot\vec{\nabla}\hat{b}$ is the curvature of the field lines and $\vec{u}_\bot$ is the velocity of the magnetic field lines. To adequately distort the magnetic flux tubes for the rate of resistive reconnection to compete with evolution requires of order $\ln (R_m)$ e-folds.  This number of e-folds in the distortion is clearly energetically impossible unless $\nu_B$ vanishes.   In three dimensions, the force exerted by the magnetic field, $\vec{f}_L$, naturally ensures that the constraint $\nu_B=0$ will be imposed. 

Three dimensionality is required to obtain the exponential enhancement of the reconnection rate by a chaotic but ideal flow of the magnetic field lines.  Nevertheless, theories that retain all three spatial coordinates may not focus on the effect of exponentiation \cite{Pontin:2012RS,Reid:2020} and instead use the presumption $d\psi_{rec}/dt \approx \eta jL$.   An extremely large current density $\approx R_m B_{rec}/\mu_0a$ is then required for reconnection to compete with evolution.  These theories omit the extreme distortions in magnetic flux tubes, which are analogous to the ``\emph{pattern milk made as it was stirred into the coffee}" that was observed by Dungey.  

The presumption that a near singular current density is required for reconnection to compete with evolution has placed a strong emphasis in three-dimensional theory on magnetic nulls.  In an ideal magnetic evolution, a singular current density arises along field lines that strike a null \cite{Craig:2014}, but the indistinguishability of magnetic field lines that come within a distance $c/\omega_{pe}$ anywhere along their trajectories makes the interpretation of this result subtle.   The separation of magnetic field lines is greatest, not least, near a null \cite{Boozer:null-X,Elder-Boozer}.

Three dimensionality, even on the small spatial scales of turbulence, enhances the reconnection rate by making the magnetic field lines chaotic.  Early work on this topic was the 1999 paper by Lazarian and Vishniac \cite{Lazarian:1999}, and work continues on this area to the present \cite{Eyink:2011,Eyink:2015,Matthaeus:2015,Lazarian:2020rev,Matthaeus:2020}.  But as discussed in the Introduction, large-scale stirring enhances topology breaking over large spatial scales with much smaller flows than those required by turbulence.  For system-wide mixing, the optimal spatial scale for the flows is system-wide.

As stated in the Introduction, the objective of this paper is to help readers obtain an understanding of the importance and nature of methods based on Lagrangian coordinates.  The concept of a Lagrangian analysis arose in the eighteenth century.  Sir Horace Lamb in Section 3 of Chapter I of \emph{Hydrodynamics} discusses \cite{Lamb:1932} the Eulerian and Lagrangian analysis of fluid motion: ``\emph{The equations obtained in these two plans are conveniently designated, as by German mathematicians, as the `Eulerian' and the `Lagrangian' forms of the hydrodynamic equations, although both forms are due to Euler.}"  The references cited by Lamb date from 1755 to 1781.  Neither magnetic reconnection nor thermal transport can be understood without the use of Lagrangian coordinates. 

Possibly the simplest model of three-dimensional magnetic reconnection in a system with well posed boundary conditions is a pressureless plasma in a perfectly conducting cylinder of radius $a$ and height $L$ with all surfaces stationary except the top which flows with a sub-Afv\'enic velocity $\vec{v}_t=\hat{z}\times\vec{\nabla}h_t(x,y,t)$ and with an initial magnetic field that is spatially constant, $\vec{B}=B_0\hat{z}$.  The stream function $h_t$ can be taken to represent the drive of coronal loops by photospheric motions.   As shown by Boozer and Elder \cite{Boozer-Elder}, even with the weakest spatial dependence consistent with the flow driving the system only in the $r<a$ regions and a simple time dependence, streamlines in the top surface separate exponentially in time and within a few evolution times bring the field into a state in which reconnection is inevitable.  The current density in the plasma is lies in ribbons that are thin but have a great width with even the sign of the current changing over short distances.  Much more can be learned about the general reconnection problem by studying even the simplest of reconnection models.

In summary, an analysis based on Lagrangian coordinates shows that traditional reconnection theories contain four false presumptions.
\begin{enumerate}
\item Two-dimensional analyses do not provide a reliable guide for magnetic reconnection in three-dimensional space.

\item Rapid  changes in magnetic topology do not require that the non-ideal part of the electric field $\mathcal{E}_{ni}$ equal $\big| u_\bot \times \vec{B}\big|$. 


\item Plasma inertia does not determine the onset rapid topological changes.


Once reconnection occurs inertia does enter through Alv\'en waves.  Information about the changed state of the magnetic field propagates across the field as compressional Alfv\'en waves and along the field as shear Alfv\'en waves.  Although Alfv\'en waves are consistent with an ideal magnetic evolution, they can drive enhanced distortions in the tubes of magnetic flux, which can produce additional reconnection.  An ideal instability of the magnetic field can also cause the evolution time to become Alfv\'enic.  

Hall terms, which produce an electric field perpendicular to the magnetic field, can affect reconnection when the evolution time is comparable to inertial time scales, Alfv\'enic or sonic, but have no direct effect when the evolution time is far slower.

\item The current density does not have an exponential increase in time although the non-ideal part of the the magnetic field does.

Current sheets naturally form, but the current density, unlike the non-ideal part of the magnetic field does not have an exponential increase with time.  

Traditional analyses often start with a sufficiently large current density in a sheet to produce rapid reconnection, $j\propto R_m$, and do not show that such a current density naturally arises in an evolution from an initial magnetic field with a current density $j \lesssim B_{rec}/\mu_0L$ with $B_{rec}$ the part of the magnetic field that is reconnecting.

\end{enumerate}



\section*{Acknowledgements}
This work was supported by the U.S. Department of Energy, Office of Science, Office of Fusion Energy Sciences under Award Numbers DE-FG02-95ER54333, DE-FG02-03ER54696, DE-SC0018424, and DE-SC0019479.



\appendix

 
 \section{Evolution of $\vec{\nabla}\cdot\vec{v}$ and $\vec{\nabla}\times\vec{v}$ \label{sec:v-ev} }
 
The equation of motion of an ideal gas subject to the force of gravity, Equation (\ref{fluid-force}), illustrates the fundamentally different evolution  of $\vec{\nabla}\cdot\vec{v}$ from $\vec{\nabla}\times\vec{v}$.

When $\vec{\nabla}\cdot\vec{v}\neq0$, Equation (\ref{fluid-force}) for the velocity evolution can be linearized to $\partial \vec{v}/\partial t = -(\vec{\nabla}p)/\rho + \vec{g}$ while studying the evolution of a velocity divergence; the pressure is $p=\rho T/m_p$.
\begin{eqnarray}
\frac{\partial (\vec{\nabla}\cdot\vec{v})}{\partial t}  &=& -\vec{\nabla}\cdot\left(\frac{\vec{\nabla}p}{\rho} -  \vec{g}\right) \\
&=&-\vec{\nabla}\cdot\left(\frac{T}{m_p} \frac{\vec{\nabla}\rho}{\rho} + \frac{\vec{\nabla}T}{m_p}\right).
\end{eqnarray} 
When the velocity is small, the density and temperature are close to their $\vec{v}=0$ values, $\rho = \rho_0 + \tilde{\rho}$ and $T=T_0 +\tilde{T}$, where $\partial\tilde{\rho}/\partial t=-\rho_0 \vec{\nabla}\cdot\vec{v}$ and $\partial\tilde{T}/\partial t=-(2T_0/3) \vec{\nabla}\cdot\vec{v}$.  Using these relations
\begin{eqnarray}
\frac{\partial^2 (\vec{\nabla}\cdot\vec{v})}{\partial t^2}  &=& \frac{5T_0}{3m_p} \nabla^2(\vec{\nabla}\cdot\vec{v}),
\end{eqnarray} 
so a divergence in the velocity, $\vec{\nabla}\cdot\vec{v}$, propagates as a sound wave through an ideal gas.


When $\vec{\nabla}\cdot\vec{v}=0$ but $\vec{\nabla}\times\vec{v}\neq0$, a vector identity implies $\vec{v}\cdot\vec{\nabla}\vec{v}=\vec{\nabla}v^2/2 -\vec{v}\times(\vec{\nabla}\times\vec{v})$.  The curl of Equation (\ref{fluid-force}) for the velocity evolution implies that $\vec{\omega}\equiv\vec{\nabla}\times\vec{v}$ obeys
\begin{eqnarray}
\frac{\partial \vec{\omega} }{\partial t}+ \vec{v}\cdot \vec{\nabla}\vec{\omega} - \vec{\omega}\cdot \vec{\nabla}\vec{v} &=& -\vec{\nabla}\times\left(\frac{\vec{\nabla}p}{\rho}\right) \\
&=&  -\vec{\nabla}\times\left(\frac{T\vec{\nabla}p}{m_pp}\right) \nonumber\\
&\simeq& \frac{\vec{g} \times \vec{\nabla}T }{T}. \label{vorticity ev}
\end{eqnarray}
using $p=\rho T/m_p$ and $\vec{\nabla}p \simeq\rho \vec{g}$. When $\hat{z}$ is a symmetry direction, so the flow is in the $x-y$ plane $\vec{\omega}\cdot \vec{\nabla}\vec{v}=0$, and $\vec{v} =-\vec{\nabla}\times(h_v\hat{z})$ so $v_x=-\partial h_v/\partial y$, $v_y=\partial h_v/\partial x$, and $\omega=\nabla^2h_v(x,y,t)$.  The gravitational acceleration is $\vec{g}=-g\hat{y}$ and $\vec{g} \times \vec{\nabla}\tilde{T} /T_0 = \hat{z} g (\partial \tilde{T}/\partial x)/T_0$, assuming the spatially variable part of the temperature $\tilde{T}$ is small compared to the spatially averaged temperature $T_0$.  Equation (\ref{vorticity ev}) then produces a vorticity that is non-zero but has a zero spatial average; $\Big< (g/T)(\partial \tilde{T}/\partial x)\Big>=0$,  so  $d\Big< \omega\Big>/dt=0$.




\end{document}